\documentclass[aip,pof,preprint]{revtex4-1}
\usepackage{amsmath}
\usepackage{amssymb}
\usepackage{graphicx}
\usepackage{dcolumn}
\usepackage{bm}

\usepackage{graphicx}
\usepackage{dcolumn}
\usepackage{bm}
\usepackage{ulem}
\usepackage{docs}%

\expandafter\ifx\csname package@font\endcsname\relax\else
 \expandafter\expandafter
 \expandafter\usepackage
 \expandafter\expandafter
 \expandafter{\csname package@font\endcsname}%
\fi
\providecommand{\ve}[1]{\mbox{\boldmath $#1$}}

\newcommand{\fp}[2]{\frac{\partial #1}{\partial #2}}

\newcommand{\boldomega}{{\boldsymbol\omega}}
\newcommand{\boldepsilon}{{\boldsymbol\epsilon}}
\newcommand{\boldsigma}{{\boldsymbol\sigma}}

\begin{document}

\author{P.-T. BRUN}
\email{brun@lmm.jussieu.fr}

\affiliation{CNRS and UPMC Univ.~Paris 06, UMR 7190, Institut Jean le Rond 
d'Alembert, Paris, France}

\affiliation{Laboratoire FAST, UPMC-Paris 6, Universit\'e Paris-Sud, CNRS, 
B\^atiment 502, Campus Universitaire, Orsay 91405, France}

\author{N. M. RIBE}
\affiliation{Laboratoire FAST, UPMC-Paris 6, Universit\'e Paris-Sud, CNRS, 
B\^atiment 502, Campus Universitaire, Orsay 91405, France}

\author{B. AUDOLY}
\affiliation{CNRS and UPMC Univ.~Paris 06, UMR 7190, Institut Jean le Rond 
d'Alembert, Paris, France}

\title{A numerical investigation of the fluid mechanical sewing machine }

\begin{abstract}

A thin thread of viscous fluid falling onto a moving belt generates a
surprising variety of patterns depending on the belt speed, fall
height, flow rate, and fluid properties.  Here we simulate this
experiment numerically using the Discrete Viscous Threads method that
can predict the non-steady dynamics of thin viscous filaments,
capturing the combined effects of inertia, stretching, bending and
twisting.  Our simulations successfully reproduce nine out of ten
different patterns previously seen in the laboratory, and agree
closely with the experimental phase diagram of Morris et al.\ (2008).
We propose a new classification of the patterns based on the Fourier
spectra of the longitudinal and transverse motion of the point of
contact of the thread with the belt.  These frequencies appear to be
locked in most cases to simple ratios of the frequency $\Omega_c$ of
steady coiling obtained in the limit of zero belt speed.  In particular
the intriguing `alternating loops' pattern is produced by combining
the first five multiples of $\Omega_c/3$.

\end{abstract}

\date{October 2011}%

\maketitle

\tableofcontents

\section{Instabilities of viscous threads}

\subsection{Introduction}

A thin stream or jet of liquid falling onto a fixed surface is one of
the simplest situations in fluid mechanics, yet it can generate a
remarkable range of phenomena.  Fast jets produce hydraulic jumps,
which are circular when the viscosity is very low \cite{Rayleigh:1914p5032,Watson:1964p4931} and
polygonal when it is somewhat higher \cite{ellegaard98}.  Thin streams
of very viscous fluid can exhibit steady coiling \cite{barnes58} or
rotatory folding \cite{habibi10}, and under some conditions coiling
produces spiral waves of air bubbles in the thin fluid layer spreading
over the surface \cite{habibi08}.  Finally, thin streams of
non-Newtonian fluid can exhibit the Kaye ("leaping shampoo'') effect
in which the stream rebounds episodically from the pile of previously
deposited fluid \cite{kaye63}.

A further degree of complexity is introduced if the source of the jet
and the surface onto which it falls are in relative motion.  This is
the case when a home cook lays down `squiggles' of icing or molten
chocolate on a cake, or when an artist lets paint dribble onto a
canvas from a moving paintbrush, a technique used to great effect by
Jackson Pollock \cite{Maha-paint}.  An analogous situation involving
many interacting jets is the `spunbonding' process of non-woven fabric
production, in which thousands of threads of molten polymer solidify
and become entangled as they fall onto a moving belt, producing a
fabric with a random texture.

\begin{figure*}
    \centerline{\includegraphics{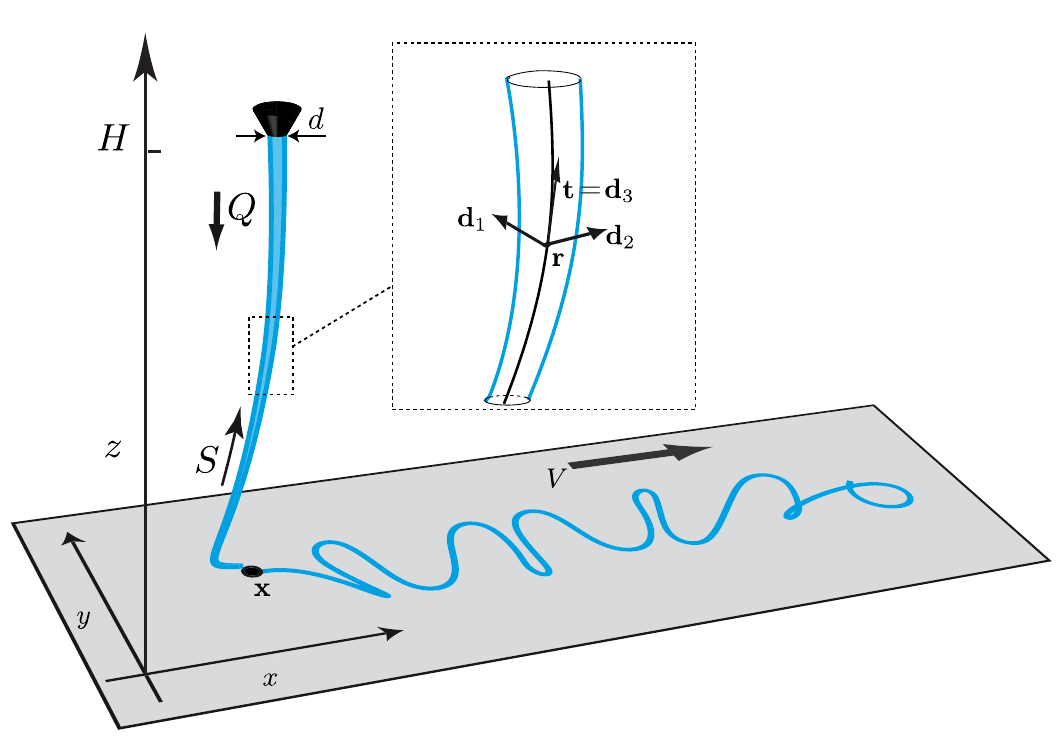}}
    \caption{Configuration of the fluid mechanical sewing machine. Newtonian
    fluid with constant density $\rho$, viscosity $\nu$, and surface
    tension coefficient $\gamma$ is ejected at a volumetric rate $Q$
    through a nozzle of diameter $d$ at a height $H$ above a belt
    moving in its own plane at constant speed $V$.  The position of
    the contact point between the thread and the belt is
    $\mathbf{x}(t)$.  Lateral advection of the contact point motion at
    speed $V$ creates complex `stitch' patterns on the belt.  In
    inset: geometry of an element of the thread, showing the
    orthonormal triad of basis vectors $\mathbf{d}_1$, $\mathbf{d}_2$,
    and $\mathbf{d}_3$ as a function of the Lagrangian coordinate $S$
    along the center-line.\label{fig:belt}}
\end{figure*}

Here we use a numerical approach to study an idealized model for these
processes: the continuous fall of a single thread of viscous fluid
onto a belt moving with a constant velocity in its own plane (Figure~\ref{fig:belt}). 
This system was first studied experimentally by
Chiu-Webster and Lister \cite{ChiuWebster:2006p1764} (henceforth CWL),
who called it the `fluid mechanical sewing machine' on account of the
stitch-like patterns traced on the belt by the thread.  The complexity
and diversity of these patterns testifies to a rich nonlinear dynamics
and bifurcation structure.  The appeal of the system is further
increased by the theoretical and numerical challenges involved in
modeling it.

In CWL's experiments, viscous fluid (corn syrup) with density $\rho$,
surface tension coefficient $\gamma$ and viscosity $\nu$ was ejected
at a volumetric rate $Q$ from a vertical nozzle of diameter $d$,
from which it fell a distance $H$ onto a belt moving at speed $V$ (fig.
\ref{fig:belt}).  The experiments were conducted by varying $V$ and
$H$ for several different combinations of values of $d$, $Q$ and
$\nu$.  When $V$ greatly exceeded a fall height-dependent critical
value $V_b(H)$, the fluid thread had the form of a steady dragged
catenary.  As the belt speed was decreased towards $V_b$, the
lowermost part of the thread evolved into a backward-facing `heel',
which became unstable to periodic meandering when $V=V_b$.  Further
decrease of the belt speed led to a series of bifurcations to more
complex patterns (fig.~\ref{fig:patterns}), ending with the
establishment of steady coiling for $V=0$.  CWL successfully predicted
the shape of the steady dragged catenary using a `viscous string'
model that neglected bending stresses.  However, this solution ceases
to exist when the extensional axial stress at the bottom of the thread
becomes zero, corresponding to the incipient formation of a heel in
which the axial stress is compressional.  Because a state of axial
compressive stress is a necessary condition for the buckling of a
slender body \cite{taylor68}, CWL interpreted the onset of meandering
as a buckling instability of the heel.

Ribe et al.~\cite{Ribe:2006p692} carried out a numerical linear
stability analysis of the dragged catenary state to predict the
critical belt speed $V_b$ and the frequency $\omega_b$ for the onset
of meandering, using a more complete `viscous rod' theory
incorporating bending and twisting of the filament.  The numerical
predictions of $V_b$ and $\omega_b $ thereby obtained agree closely
with the experimental measurements of \cite{ChiuWebster:2006p1764}.
Ribe et al.~\cite{Ribe:2006p692} also documented a close
correspondence between incipient meandering and finite-amplitude
steady coiling on a motionless ($V=0$) surface, such that $\omega_b$
is nearly identical to the steady coiling frequency $\Omega_c$ for any
given fall height $H$.  Moreover, the critical belt speed $V_b(H)$ is
nearly identical to the vertical (free-fall) speed $U_f$ of the fluid
at the bottom of the thread, indicating that meandering sets in when
the belt is no longer moving fast enough to carry away in a straight
line the fluid falling onto it.

More extensive experiments were conducted by Morris et al.
\cite{Morris:2008p1754}, using a carefully engineered apparatus with
silicone oil as the working fluid for better stability and
reproducibility.  They determined a complete phase diagram for the
patterns as a function of $H$ and $V$ for a particular set of values
of the fluid viscosity $\nu$ and the flow rate $Q$.  They showed that
the observed amplitude of meandering as a function of the belt speed
is consistent with a Hopf bifurcation, and proposed a simple model to
predict it based on the hypothesis that the contact point moves at
constant speed relative to the belt.  Finally, they proposed a generic
set of amplitude equations which they used to characterize the
alternating loops (which they called `figure-of-eight') and translated coiling patterns. \\

Most recently, Blount and Lister \cite{blount11} performed a detailed
asymptotic analysis of a slender dragged viscous thread, focussing on
the structure of the heel.  They showed that the lowermost part of the
thread can exhibit three distinct dynamical regimes depending on
whether the belt speed is greater than, nearly equal to, or less than
the free-fall speed $U_f$.  Their asymptotic stability analysis of
these steady states indicates that meandering sets in when the
horizontal reaction force at the belt begins to be slightly against
the direction of belt motion, corresponding to the heel `losing its
balance'.

As the above summary indicates, our current theoretical understanding
of the fluid-mechanical sewing machine is essentially limited to the
initial bifurcation from the steady dragged configuration to
meandering.  In this paper, we push forward into the fully nonlinear
regime with the help of a new computational
algorithm~\cite{%
Bergou-Audoly-EtAl-Discrete-Viscous-Threads-2010,%
Audoly-Clauvelin-EtAl-Simulating-the-dynamics-of-thin-2011%
}
that permits for the first time robust numerical modeling of arbitrary
non-stationary dynamics of viscous threads.  After describing the
method briefly, we use it to generate a complete phase diagram of
sewing-machine patterns that reproduces all the major features of the
diagram determined experimentally by Morris et al.
\cite{Morris:2008p1754}.  We then perform a detailed Fourier analysis
of the motion of the contact point for each of the patterns simulated,
and propose a new classification of them based on the spectral content
of the motions of the contact point in two orthogonal directions.

For most of the patterns studied, we find that the frequencies present
in the spectra of the contact point motion are multiples of the steady
coiling frequency $\Omega_c$, indicating that the dynamics of the sewing machine are closely related to those of steady coiling.  Accordingly, we set the stage for our study with a brief summary of the latter in the next section.

\subsection{Steady coiling}

\begin{figure*}
  \centerline{\includegraphics{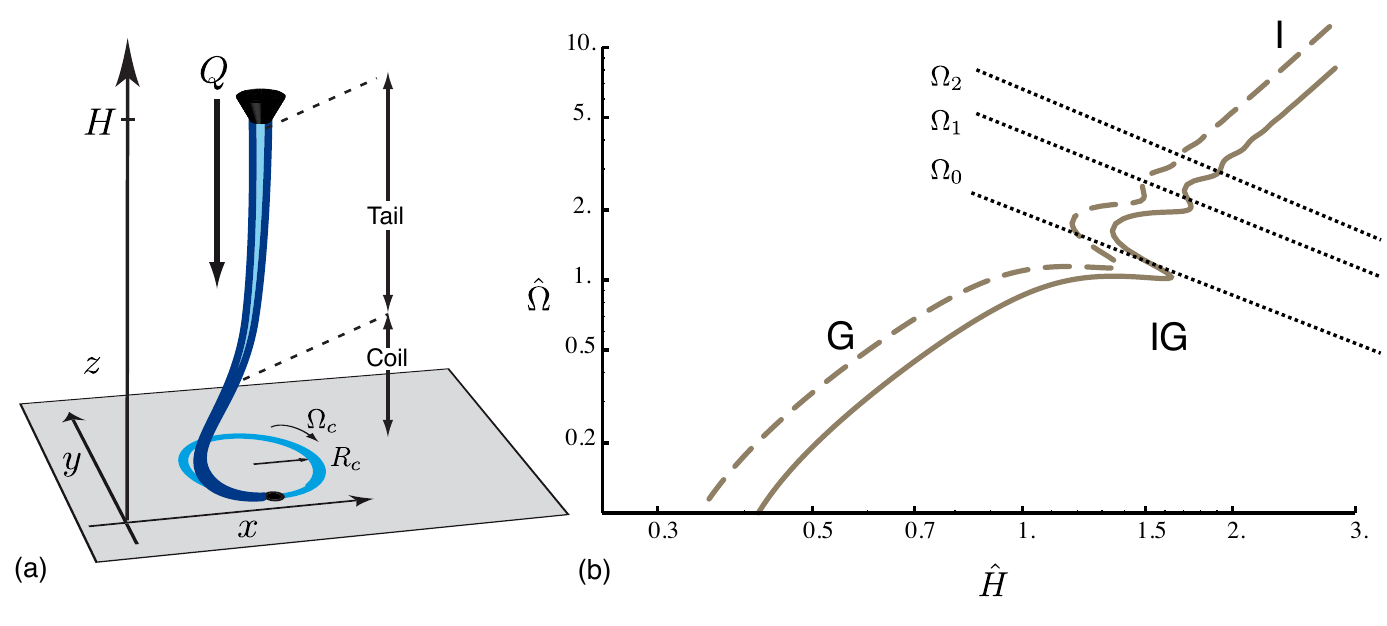}}
 \caption{
   Steady coiling of a viscous thread.  (a) Definition sketch.  (b)
   Coiling frequency as a function of height for the parameters of the
   experiments of Morris et al.~\cite{Morris:2008p1754}, calculated numerically
   using a continuation method \cite{Ribe2004}.  The solid curve
   includes the effect of surface tension ($\Pi_3 = 1.84$), while the
   dashed curve is for zero surface tension ($\Pi_3 = 0$).  The
   portions of the curves corresponding to the gravitational (G),
   inertio-gravitational (IG), and inertial (I) regimes are indicated.
   The dotted lines show the first three eigenfrequencies of a free
   viscous pendulum.  }
\label{fig:steady}
\end{figure*}

In steady coiling, the contact point of the thread with the surface
moves with a constant angular velocity $\Omega_c$ along a circle of
radius $R_c$ (Figure~\ref{fig:steady}a). 
In most cases, the thread comprises two distinct parts: a long,
roughly vertical `tail' which deforms primarily by stretching under
gravity, and a helical `coil' in which the deformation is dominated by
bending and (to a lesser extent) twisting.  Thus the radius of the
thread within the coil $a_1$ is nearly constant.  By conservation of
mass, the axial speed of the fluid in the coil is $U_1=Q/\pi a_1^2$. 

Steady coiling can occur in several distinct dynamical regimes
characterized by different balances of the viscous, gravitational, and
inertial forces acting on the thread \cite{Mahadevan:1998p1053,
Ribe2004, Ribe:2006p691}.  These regimes appear clearly on plots of
the coiling frequency vs.  the fall height.  For convenience, we
define a dimensionless fall height $\hat H$ and a dimensionless
coiling frequency $\hat\Omega$:
\begin{equation}
    \hat H = H \left(
    \frac{g}{\nu^2}
    \right)^{1/3},
    \quad
    \hat\Omega = \Omega\left(
    \frac{\nu}{g^2}
    \right)^{1/3}.
    \label{hhat}
\end{equation}
Figure~\ref{fig:steady}b shows $\hat\Omega$ as a function of $\hat H$ for the
parameters of the experiments of Morris et al.
\cite{Morris:2008p1754}, viz., $\nu = 0.0277 \,\mathrm{m}^2\,\mathrm{s}^{-1}$,
$\rho=10^3\,\mathrm{kg\,m}^{-3}$, 
$\gamma=0.0215\,\mathrm{N\,m}^{-1}$, $d= 8\,\mathrm{mm}$ and
$\rho\,Q = 0.0270\,\mathrm{g\, s}^{-1}$.	 For $\hat H < 1.2$, coiling occurs in
a gravitational (G) regime.  Inertia is negligible everywhere in the
thread, which is governed by a balance between gravity and the viscous
forces that resist stretching (in the tail) and bending (in the coil).
At intermediate heights $1.2\leq \hat H \leq 2.2$, a complex
inertio-gravitational (IG) regime appears, in which the coiling
frequency is a multivalued function of the fall height.  The
centrifugal force now becomes important in the tail, which behaves as
a distributed pendulum with an infinity of eigenmodes whose
corresponding eigenfrequencies are proportional to the simple pendulum
frequency $(g/H)^{1/2}$.  The first three of these frequencies are
shown by the dotted lines in Figure~\ref{fig:steady}b.  When one of these
eigenfrequencies is close to the frequency set by the coil, the tail
enters into resonance with the latter, giving rise to resonance peaks
that appear as rightward-facing `bumps' in the curve $\hat{\Omega}(\hat H)$.  For
large heights $\hat H > 2.2$, coiling occurs in an inertial (I) regime
in which the viscous bending force in the coil is balanced by 
inertia~\cite{Mahadevan:1998p1053}.
The tail in this regime is almost perfectly vertical, and is
controlled by a balance between gravity, the viscous stretching force,
and the axial momentum flux.  Finally, there is also a viscous (V)
regime in which both gravity and inertia are negligible everywhere in
the thread, but this is only observed when both $H$ and $d$ are much
smaller than in the experiments of CWL and Morris et al.

In a typical laboratory experiments on steady coiling, the parameters 
$d$ and $Q$ and the fluid properties $\rho$, $\nu$ and $\gamma$
are held fixed while $H$ is varied. Each such experiment is 
completely defined by the values of the three dimensionless groups
\begin{equation}
\Pi_1 = \left(\frac{\nu^5}{g\,Q^3}\right)^{1/5} ,
\quad \Pi_2 = \left(\frac{\nu\,Q}{g\,d^4}\right)^{1/4} ,
\quad \Pi_3 = \frac{\gamma \,d^2}{\rho \,\nu\, Q}
\label{pi123}
\end{equation}
As an example, $\Pi_1 = 610$, $\Pi_2 =0.370$, and $\Pi_3 =1.84$ for
all the experiments of Morris et al.~\cite{Morris:2008p1754}.  The
(dimensionless) functional dependence of the coiling frequency on the
other parameters now takes the general form
\begin{equation}
\hat\Omega=\hat{\Omega}(\hat H,\Pi_1,\Pi_2,\Pi_3). 
\end{equation}
The effect of surface tension is measured by the parameter $\Pi_3$.
Surface tension modifies the coiling frequency quantitatively but 
introduces no essentially new dynamics, as
can be seen by comparing the solid ($\Pi_3 = 1.84$) and dashed
($\Pi_3 = 0$) curves in Fig.~\ref{fig:steady}b. 

The continuation method used to generate the curves in Fig.~\ref{fig:steady}b can be used for steady coiling because the flow is
stationary in a co-rotating reference frame that moves with the
contact point.  No such reference frame exists for the sewing machine
configuration.  We therefore require a different numerical method,
which is described in the next section.

\section{Numerical method}

Our numerical experiments of the sewing machine were set up using the
computational method of Discrete Viscous Threads (henceforth DVT) originally
described in a conference
paper~\cite{Bergou-Audoly-EtAl-Discrete-Viscous-Threads-2010}, and
presented in detail in a recent
paper~\cite{Audoly-Clauvelin-EtAl-Simulating-the-dynamics-of-thin-2011}.
To the best of our knowledge, DVT is the
only available numerical method that is fast and robust enough to be
applicable to the sewing machine geometry while retaining all the
relevant modes of deformation, namely stretching, twisting and
bending.  For a thin thread the stretching modulus varies as the square of the thread's radius, while the bending and twisting moduli are proportional to the fourth power of the radius. As a result the dynamics of
thin threads is a nonlinear and numerically stiff problem.  The DVT
method addresses this difficulty by introducing an elaborate spatial
discretization of the equations based on ideas from differential
geometry.  The method allows simulations to be run for arbitrary mesh
sizes, even very coarse ones, with optimal stability.  This contrasts
with conventional discretization schemes which are typically stable
for sufficiently small mesh sizes only, the maximum mesh size being in
practice a small fraction of the smallest length scale in the flow,
here the small size of the coiled region at the bottom.

Below we briefly present the principles of the DVT method, introduce
adaptive mesh refinement which provides a tremendous speed-up of the
calculations when gravity stretches the tail significantly, validate
the code against known solutions for steady coiling, explain the
details of the numerical procedure, and finally present our numerical
results.

\subsection{Smooth case: the equations for thin viscous threads}


The numerical code makes use of the Lagrangian approach and the
viscous thread is discretized as a polygonal line.  A mass is assigned
to each vertex, forces are set up on these masses, and the motion of
each mass is obtained by time integration of the fundamental law of
dynamics.  The discrete forces are designed in such a way that the
motion of the polygonal line is equivalent to that of a thin viscous
thread in the smooth limit.  The key element of the numerical method
is the expressions for these discrete viscous forces.  To prepare the
derivation, we start by reformulating the smooth case, usually
expressed in Eulerian variables, in the Lagrangian framework.  We
introduce a Lagrangian coordinate $S$ that marks cross-sections and
follows them during motion.  This Lagrangian coordinate $S$ is defined
as the arc-length in an imaginary reference configuration where the
thread is a cylindrical tube of constant radius.  It plays the same
role as the vertex index $i$ in the discrete case.

At a particular time $t$, the configuration of the thread is defined
by its center-line $\mathbf{r}(S,t)$ and an orthonormal triad
$(\mathbf{d}_{1}(S,t),\mathbf{d}_{2}(S,t),\mathbf{d}_{3}(S,t))$.  This
triad allows one to keep track of twisting, because the directions of
$\mathbf{d}_{1}$ and $\mathbf{d}_{2}$ follow the orientation of a
cross-section as it spins about the center-line.  Thin threads deform
is such a way that cross-sections remain approximately planar and
perpendicular to the centerline.  As a result, the center-line tangent
$\mathbf{r}'(S,t) = \frac{\partial \mathbf{r}}{\partial S}$ and the
unit vector $\mathbf{d}_{3}(S,t)$ are aligned.  Denoting by
$\ell(S,t)$ the axial stretch factor based on the reference
configuration, given by the norm of $\mathbf{r}'(S,t)$, we have:
\begin{equation}
    \mathbf{r}'(S,t) = \ell(S,t)\,\mathbf{d}_{3}(S,t)
    \textrm{.}
    \label{eq:centerline}
\end{equation}

Since the triad $(\mathbf{d}_{i}(S,t))_{i=1,2,3}$ is orthonormal, its
time evolution defines a rigid-body rotation for any particular value
of $S$.  The associated instantaneous angular velocity
$\boldomega(S,t)$ is such that
\begin{equation}
    \frac{\partial \mathbf{d}_{i}(S,t)}{\partial t}
    =
    \boldomega(S,t)\times \mathbf{d}_{i}(S,t)
    \quad
    \textrm{($i=1,2,3$)}
    \textrm{.}
    \label{eq:materialRotation}
\end{equation}
One can take advantage of the fact that the vectors $\mathbf{r}'$ and
$\mathbf{d}_{3}$ must remain aligned by equation~(\ref{eq:centerline})
to capture the twisting motion of the thread using a single parameter,
the angular spinning velocity $w(S,t)$ defined by $w(S,t) = \boldomega(S,t)\cdot \mathbf{d}_{3}(S,t)$.  In
this view, which we call the centerline/spin
representation~\cite{Audoly-Clauvelin-EtAl-Simulating-the-dynamics-of-thin-2011},
the material velocity $\boldomega$ is a secondary variable which can
be reconstructed by the equation
\begin{equation}
    \boldomega(S,t) = \mathbf{d}_{3}(S,t)\times 
    \dot{\mathbf{d}}_{3}(S,t)+ w(S,t)\,\mathbf{d}_{3}(S,t)
    \textrm{.}
    \label{eq:smoothOmegaReconstruction}
\end{equation}
where  the time derivative $\frac{\partial}{\partial t}$ is denoted using a dot. In a  viscous thread, the fundamental kinematical quantities are the 
strain rates, defined by
\begin{equation}
    \dot{\epsilon}_{\mathrm{s}} 
    = \frac{1}{\ell}\,
    \frac{\partial \mathbf{u}(S,t)}{\partial S}\cdot\mathbf{d}_{3}(S,t),
    \qquad
    \dot{\boldepsilon}_{\mathrm{tb}} = \frac{1}{\ell}\,\frac{\partial 
    \boldomega(S,t)}{\partial S}
    \label{eq:smoothStrainRates}
\end{equation}
where $\mathbf{u} = \dot{\mathbf{r}}$ is the center-line velocity.
Here $\dot{\epsilon}_{\mathrm{s}}$ captures the strain rate associated
with axial stretching, while the vector
$\dot{\boldepsilon}_{\mathrm{tb}}$ captures in a combined manner the
strain rates for the twisting and bending modes.  The strain rate for
stretching, $\dot{\epsilon}_{\mathrm{s}}$, is related to the time
evolution of center-line stretch $\ell$ by the formula
$\dot{\epsilon}_{\mathrm{s}} = \partial \ln\ell/\partial t$.

For a thin thread, the internal stress is described by the resultant
$\mathbf{n}(S,t)$ of the viscous forces over a particular
cross-section, and their moment $\mathbf{m}(S,t)$ with respect to the
origin of the cross-section.  These internal force and moment vectors
play the same role as the tensor $\boldsigma$ in 3D continuum
mechanics.  For the special case of a thin thread geometry, Stokes's
constitutive law states that stress is proportional to the rate of
deformation: for the stretching mode, we have
\begin{equation}
    \mathbf{n}(S,t)\cdot \mathbf{d}_{3}(S,t) = 
    3\,\mu\,A\,
    \dot{\epsilon}_{\mathrm{s}}
    \textrm{,}
    \label{eq:constitutiveStretching}
\end{equation}
and for the bending and twisting modes,
\begin{equation}
    \mathbf{m}(S,t) = \left(
    3\mu\,I\,\mathbf{P}_{12}
    +
    2\mu\,I\,\mathbf{P}_{3}
    \right)
    \cdot
    \dot{\boldepsilon}_{\mathrm{tb}}
    \textrm{.}
    \label{eq:constitutiveBendingTwisting}
\end{equation}
Here $\mu\equiv \rho\, \nu$ is the dynamic viscosity,
$A = \pi\,a^2$ the cross-sectional area,
$I=\pi\,a^4/4$ is the moment of inertia, 
$a$ is the radius, $\mathbf{P}_{3}
= \mathbf{d}_{3}\otimes\mathbf{d}_{3}$ is the tangent projection
operator, and $\mathbf{P}_{12} = \mathbf{1} - \mathbf{P}_{3}$ is the
normal projection operator.  The radius $a(S,t)$ is a dependent
variable which is reconstructed by the incompressibility condition
\begin{equation}
    a(S,t) = \frac{a_{0}}{\sqrt{\ell(S,t)}}
    \textrm{,}
    \label{eq:radiusReconstruction}
\end{equation}
where $a_{0}=d/2$ is the radius in the configuration of reference (note
that $\ell=1$ in the reference configuration by definition).  The
stretching modulus $3\mu\,A$ was originally derived by
Trouton~\cite{Trouton-On-the-coefficient-of-viscous-traction-1906},
and the bending modulus $3\mu\,I$ can be found for instance
in~\cite{Buckmaster-Nachman-EtAl-The-buckling-and-stretching-of-a-viscida-1975}.

These equations are complemented by the balance of linear and angular
momentum, known as the Kirchhoff equations for thin rods:
\begin{align}
   \fp{\mathbf{n}(S,t)}{S} + \mathbf{P}(S,t)  & = 
    \rho\,A_{0}\,    \fp{{}^2 \mathbf{r}(S,t)}{t^2}
    \label{eq:KirchhoffEqnsLagrangian-Transl} \\
    \fp{\mathbf{m}(S,t)}{S} + \frac{\partial \mathbf{r}(S,t)}{\partial
    S}\times \mathbf{n}(S,t) & = \mathbf{0}
    \textrm{.}
    \label{eq:KirchhoffEqnsLagrangian-Rot}
\end{align}
Following an approximation introduced by Kirchhoff himself which is
valid for thin threads, we have neglected the rotational inertia in
the second equation.  The vector $\mathbf{P}(S,t)$ is the resultant of
the external forces per unit \textit{reference} length $\mathrm{d}S$.
The weight of the thread and the surface tension are taken into
account by setting
\begin{equation}
    \mathbf{P} = \rho\,A_{0}\,\mathbf{g} - \frac{\partial 
    (\pi\,\gamma\,a(S,t)\,\mathbf{d}_{3}(S,t))}{\partial S}
    \label{eq:smoothExternalForces}
\end{equation}
where $A_{0} = \pi\,{a_{0}}^2$ is the cross-sectional area in the
reference configuration, $\mathbf{g}$ the acceleration of gravity, and
$\gamma$ is the surface tension at the fluid-air interface.  The last
term in equation~(\ref{eq:smoothExternalForces}) is the net force on
the center-line due to surface tension at the fluid-air interface, the
argument in the derivative being the axial force due to the capillary
overpressure $(\pi\,a^2)\,(\frac{\gamma}{a})$.  Note that there is no need
to consider a linear density of applied torque in
equation~(\ref{eq:KirchhoffEqnsLagrangian-Rot}) for the problem at
hand.

With suitable boundary conditions, the set of partial differential
equations~(\ref{eq:centerline}--\ref{eq:KirchhoffEqnsLagrangian-Rot})
constitutes a well-posed mathematical problem governing the dynamics
of a viscous thread.

\subsection{A variational view: Rayleigh potentials}

The equations of
motion~(\ref{eq:KirchhoffEqnsLagrangian-Transl}--\ref{eq:KirchhoffEqnsLagrangian-Rot})
and the constitutive
law~(\ref{eq:constitutiveStretching}--\ref{eq:constitutiveBendingTwisting})
can be discretized in a natural
manner~\cite{Audoly-Clauvelin-EtAl-Simulating-the-dynamics-of-thin-2011}
if they are first re-written in terms of a Rayleigh potential.  The
Rayleigh potential $\mathcal{D}$ yields the power dissipated by
viscous forces as a function of the vertex velocity $\mathbf{u}(S) =
\dot{\mathbf{r}}(S)$ and spinning velocity $w(S) = \boldomega(S)\cdot
\mathbf{d}_{3}(S)$.  For a thin thread, it has three contributions
corresponding to the stretching, bending and twisting modes,
$\mathcal{D}(\mathbf{u},w) = \mathcal{D}_{\mathrm{s}}(\mathbf{u}) +
\mathcal{D}_{\mathrm{t}}(\mathbf{u},w) +
\mathcal{D}_{\mathrm{b}}(\mathbf{u},w)$.  Note that the Rayleigh
potential $\mathcal{D}$ also depends on the current configuration
$\mathbf{r}(S,t)$ but this dependence will be implicit in our
notations.  As an illustration, consider the stretching contribution
$\mathcal{D}_{\mathrm{s}}$.  It only depends on the vertex velocities,
and reads
\begin{equation}
    \mathcal{D}_{\mathrm{s}}(\mathbf{u}) = \frac{1}{2}\,\int 
    3\,\mu\,A\,(\dot{\epsilon}_{\mathrm{s}})^2\,\ell\,\mathrm{d}S
    \label{eq:smoothStretchingDissipationPotential}
\end{equation}
where $\dot{\epsilon}_{\mathrm{s}}$ in the integrand is given by
equation~(\ref{eq:smoothStrainRates}), and $\ell\,\mathrm{d}S$ is the
infinitesimal arc-length in current configuration.  The expressions
for the twist and bending contributions can be found in
reference~\cite{Audoly-Clauvelin-EtAl-Simulating-the-dynamics-of-thin-2011}.

The Rayleigh dissipation potential is useful as it captures the effect
of the internal viscous stress in a compact mathematical form.  Indeed
in the equations of
motion~(\ref{eq:KirchhoffEqnsLagrangian-Transl}--\ref{eq:KirchhoffEqnsLagrangian-Rot})
the net viscous force $\mathbf{n}'$ and the
net viscous moment $\mathbf{m}' + \mathbf{r}'\times \mathbf{n}$ in the
left-hand sides can be shown to be equivalent to a density of applied
force
\begin{equation}
    \mathbf{P}_{\mathrm{v}}(S,t) = -
    \left.
    \frac{\partial \mathcal{D}(\hat{\mathbf{u}},\hat{w})}{\partial 
    \hat{\mathbf{u}}(S)}
    \right|_{(\hat{\mathbf{u}} = \dot{\mathbf{x}}(t),\hat{w}=w(t))}
    \label{eq:PvSmooth}
\end{equation}
and a density of applied twisting torque
\begin{equation}
    Q_{\mathrm{v}}(S,t) = -
    \left.
    \frac{\partial \mathcal{D}(\hat{\mathbf{u}},\hat{w})}{\partial 
    \hat{w}(S)}
    \right|_{(\hat{\mathbf{u}} = \dot{\mathbf{x}}(t),\hat{w}=w(t))}
    \textrm{.}
    \label{eq:QvSmooth}
\end{equation}
In these equations, the right-hand sides denote \textit{functional}
derivatives, as the dissipation potential $\mathcal{D}$ takes the
functions $\hat{\mathbf{u}}(S)$ and $\hat{w}(S)$ as arguments.  The
hat notation expresses the fact that the derivatives have to be
calculated formally first, and evaluated with using the real motion
$(\mathbf{u} = \dot{\mathbf{r}}, w)$ next.

\subsection{Discretization}

In the discrete case, the center-line of the thread is represented by
a polygonal chain of $n+2$ particles $\mathbf{R}(t) = \{
\mathbf{r}_0(t), \mathbf{r}_{1}(t), \cdots, \mathbf{r}_{n+1}(t)\}$.
The length $\ell^i(t)$ and unit tangent $\mathbf{d}_{3}^i(t)$ of an
edge $i$ are defined by
\begin{equation}
    \mathbf{r}_{i+1}(t) - \mathbf{r}_{i}(t) = \ell^i(t)\,
    \mathbf{d}_{3}^i(t)
    \textrm{.}
\end{equation}

We consider viscous threads having a circular cross-section.  As a
result there is no need to keep track of the absolute orientation of
the cross-sections during motion.  Twist is taken into account through
the instantaneous angular velocity of spin of an edge, noted $w^i(t)$.
This is an unknown of the motion, for which we will derive an
equation.  Representing rotations by a single degree of freedom
is beneficial for the simulation.  The angular velocity vector
$\boldomega^i$ is a secondary quantity in the simulation, which is
reconstructed from the spinning velocity $w^i$ by an equation similar
to the smooth equation~(\ref{eq:smoothOmegaReconstruction}).

The generalized velocity of a viscous thread is obtained by
complementing the vertex velocities
$\dot{\mathbf{R}}(t)=\{\dot{\mathbf{r}_{0}}(t),\cdots\}$ with the
spinning velocities of the edges $\mathbf{W}(t) = \{w^0(t),\cdots\}$.
The dynamics of the thread is specified by a differential equation
involving the position $\mathbf{R}(t)$, velocities $\dot{\mathbf{R}}(t)$,
$\mathbf{W}(t)$, as well as the acceleration $\ddot{\mathbf{R}}(t)$.
Rotational inertia is neglected and so $\dot{\mathbf{W}}(t)$ does not
enter into the equation.  This differential equation is derived next.

As in the smooth case, internal viscous stress is captured by means of
a discrete Rayleigh dissipation potential which is the sum of three
contributions, $\mathcal{D}(\mathbf{U},\mathbf{W}) =
\mathcal{D}_{\mathrm{s}}(\mathbf{U}) +
\mathcal{D}_{\mathrm{t}}(\mathbf{U},\mathbf{W}) +
\mathcal{D}_{\mathrm{b}}(\mathbf{U},\mathbf{W}) $.  As an
illustration, the stretching contribution is defined in close analogy
with equation~(\ref{eq:smoothStretchingDissipationPotential}) by
\begin{equation}
    \mathcal{D}_{\mathrm{s}}(\mathbf{U}) = \frac{1}{2}\,\sum
    D^i\,(\dot{\epsilon}_{\mathrm{s}}^i)^2
    \label{eq:discreteStretchingDissipationPotential}
\end{equation}
where $D^i = 3\,\mu\,A^i\,\ell^i$ is a discrete stretching modulus
defined by analogy with equation~(\ref{eq:constitutiveStretching}),
and $\dot{\epsilon}_{\mathrm{s}}^i =
\frac{1}{\ell^i}\,\mathbf{d}_{3}^i\cdot(\mathbf{u}_{i+1} -
\mathbf{u}_{i})$ is a discrete axial strain rate defined by analogy
with equation~(\ref{eq:smoothStrainRates}).  For an expression of the
twist and bending contributions $\mathcal{D}_{\mathrm{t}}$ and
$\mathcal{D}_{\mathrm{b}}$, we refer the reader
to~\cite{Audoly-Clauvelin-EtAl-Simulating-the-dynamics-of-thin-2011}.
They make use of discrete notions of curvature and twist, based on
ideas from discrete differential geometry.

In analogy with the smooth case, we write the equations of motion as
\begin{align}
    \mathbf{P}_{\mathrm{v}}(t)
    + \mathbf{P}(t)
    & = 
    \mathbf{M}\cdot \ddot{\mathbf{R}}(t)
    \label{eq:discreteEqnsOfMotion-vertex}\\
    \mathbf{Q}_{\mathrm{v}}(t)
    & =
    \mathbf{0}
    \label{eq:discreteEqnsOfMotion-edge}
    \textrm{.}
\end{align}
The first equation is a balance of linear momentum for the vertices
and is associated with the positional degrees of freedom
$\mathbf{R}(t)$, while the second equation is a balance of twisting
torque at each edge, associated with the spinning degrees of freedom
$\mathbf{W}(t)$.  Here, $\mathbf{M}$ is the diagonal matrix filled
with the mass of the vertices, $\mathbf{P}_{\mathrm{v}}$ and
$\mathbf{Q}_{\mathrm{v}}$ are the viscous forces and twisting
torques representing the internal stress in the thread, and
$\mathbf{P}$ combines the weight and the net force on vertices due to
surface tension.  For a detailed derivation of the discrete surface
tension forces,
see~\cite{Audoly-Clauvelin-EtAl-Simulating-the-dynamics-of-thin-2011}.
As in equation~(\ref{eq:KirchhoffEqnsLagrangian-Rot}) for the smooth
case, we have neglected rotational inertia in the right-hand side of
equation~(\ref{eq:discreteEqnsOfMotion-edge}): we have checked that
this has negligible effect on the simulation when the thread is thin
enough.

Our discretization of the thread is based on the Rayleigh potentials,
and the discrete viscous forces and moments are defined as in 
equations~(\ref{eq:PvSmooth}) and~(\ref{eq:QvSmooth}) by:
\begin{align}
    \mathbf{P}_{\mathrm{v}}(t) & = -
    \left.\frac{\partial \mathcal{D}(
    \hat{\mathbf{U}},\hat{\mathbf{W}}
    )}{\partial 
    \hat{\mathbf{U}}}
    \right|_{(\hat{\mathbf{U}} = \dot{\mathbf{R}}(t),\hat{\mathbf{W}} 
    = \mathbf{W}(t))}\\
    \mathbf{Q}_{\mathrm{v}}(t) & = -
    \left.\frac{\partial \mathcal{D}(
    \mathbf{U},\mathbf{W}
    )}{\partial 
    \mathbf{W}}
    \right|_{(\hat{\mathbf{U}} = \dot{\mathbf{R}}(t),\hat{\mathbf{W}}
    = \mathbf{W}(t))}
    \textrm{.}
\end{align}

The equations of the present section provide a complete system of
equations for the dynamics of a discrete viscous thread.  For the time
discretization, we use a semi-implicit Euler scheme, which provides
good stability even for quite large time-steps (by semi-implicit, we
mean that we linearize the equations near the current configuration at
every time-step, before applying an implicit scheme).  The treatment
of boundary conditions is explained in
section~\ref{ssec:boundaryConditions}.

\subsection{Adaptive mesh refinement}
\label{ssec:adaptivity}

The DVT method uses a Lagrangian grid which is advected by the flow.
In sewing machine experiments, gravity can typically stretch the
centerline by a factor 5 to 10 over the course of the descent.  In the
absence of refinement, this makes the grid very inhomogeneous: to
capture the small-scale features near the belt one has to use an
extremely fine mesh size near the nozzle.  Overall, a large number of
degrees of freedom are required and the simulation is inefficient.  In
addition, important inhomogeneities in edge lengths makes the
time-stepping problem ill-conditioned and robustness is affected.  To
overcome these difficulties, we have set up adaptive mesh
refinement.

In our implementation of refinement, edges are subdivided whenever
their length exceeds a nominal length, which is a prescribed function
of the distance to the belt.  In the upper part of the belt, this
nominal length is equal to twice the initial segment length, defined
by the periodic release of new (Lagrangian) vertices from the nozzle.
To better resolve the coil region, this nominal segment length
decreased near the belt according to a prescribed exponential profile.
This profile was adjusted in such a way that the coil region always
includes a sufficient number of vertices, typically 10 to 30, with a
final edge length typically below $0.006\, (\nu^2/g)^{1/3}$, and that
the interval between subsequent subdivisions of a given edge is always
larger than two simulation steps.

Whenever an edge was marked as needing subdivision, a new vertex was
inserted.  We computed the properties of the new vertex and of the two
new edges as follows: the Lagrangian coordinate of the new vertex is
obtained by linear interpolation, the mass stored in the former edge
is equally split among its children, the position and velocity of the
new vertex are calculated by an interpolation of order 4, the
cross-sectional area $A$, the spinning velocities, and the viscosities
of the new edges are obtained by second-order interpolation.  These
interpolation orders were chosen in such a way that the bending and
twist forces remain continuous upon subdivision.  We used the steady
coiling geometry to adjust the subdivision parameters and to validate
the subdivision scheme.

\subsection{Emission from the nozzle, capture by the belt}
\label{ssec:boundaryConditions}

We found that the implementation of boundary conditions was a key
point to successfully reproduce the patterns and the phase diagram of
the experimental sewing machine.  We tried simple implementations
first, and could reproduce the curves $\hat{\Omega}_{c}(\hat{H})$ for
the frequency of steady coiling as well as the simplest stitch
patterns, but failed to reproduce entire regions of the phase
diagram.  Further examination revealed the presence of spurious
oscillations in the calculated acceleration in the steady coiling
geometry ($\hat{V}=0$), even though the coiling frequency
$\hat{\Omega}_{c}(\hat{H})$ was correctly predicted.  We removed these
spurious oscillations by a more careful account of the boundary
conditions both at the nozzle and at the belt, as explained below.
Suppressing these oscillations appeared to be sufficient to bring the
numerical predictions in close agreement with the experimental ones.

New vertices need to be emitted periodically from the nozzle.  In a
first implementation of the clamped boundary conditions there, we
considered an infinite string of fluid particles which were moved with
the prescribed ejection velocity $Q/A_{0}$, until they passed below
the nozzle and were released.  The position of the first vertex
clamped inside the nozzle varies abruptly as a new vertex is released,
and this was the cause of unwanted oscillations.  They were suppressed
by considering a string of two particles in the nozzle, with a fixed
position relative to the nozzle; the injection velocity is then
modelled by steadily increasing the length of the second edge, and
periodically inserting a new vertex in third position.

Impact with the belt was first handled by detecting penetration of
vertices into the belt at the end of every time-step, and constraining
their velocity to match the belt's velocity at any subsequent time.
This also induces large unwanted fluctuations in the acceleration,
which can be interpreted by the fact that the vertical momentum
resulting from the collision is not transferred to the thread until
the following time step.  The oscillations were removed by using a
technique known as roll-back.  Then, every time-step involves an
iteration whose aim is to determine the set of vertices undergoing a
collision during the time-step: whenever an unexpected collision takes
place, the step is discarded, time is rolled back, and a new time-step
is computed, forcing additional vertices to land on the belt at the
end of the time-step.  An additional difficulty in the implementation
of roll-back in the context of a linearized implicit scheme, is that a
only small number of particles can be captured at every step.  We
circumvented this difficulty by using adaptive time refinement.  Such
an adaptive time refinement is presumably not needed if a fully
(non-linear) implicit scheme is used, such as the one presented in
Ref.~\cite{Bergou-Audoly-EtAl-Discrete-Viscous-Threads-2010}.

\subsection{Validation}

\begin{figure}
  \centerline{\includegraphics[%
width=.6\textwidth
]{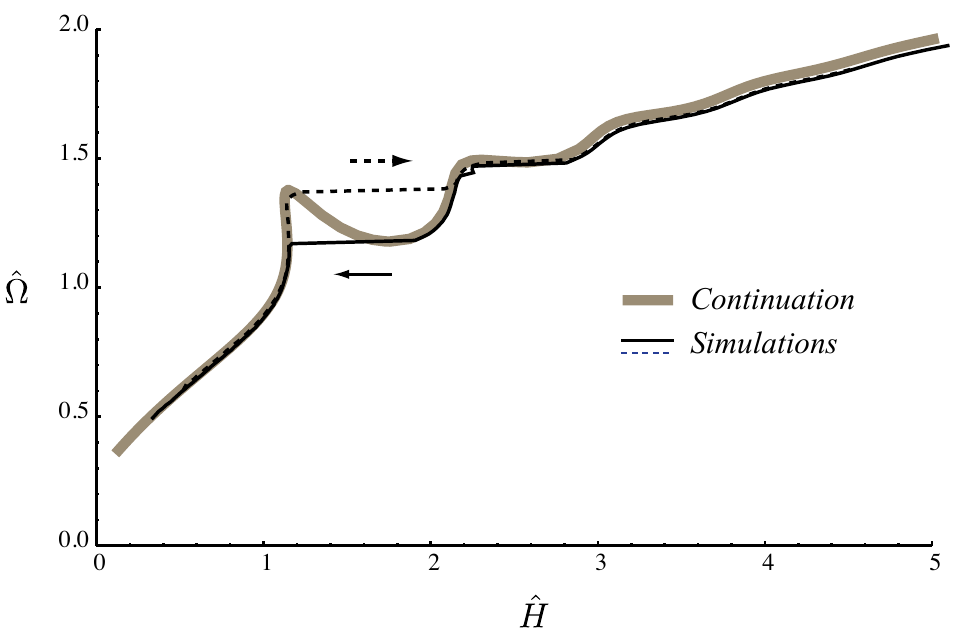}}
  \caption{Validation of the discrete numerical algorithm for a
  steadily coiling viscous thread with $\Pi_1 = 610$, $\Pi_2 = 0.37$,
  and $\Pi_3 = 0$.  The discrete simulations with the fall height
  increasing (dashed line) and decreasing (solid line) match closely
  the solution of the steady-state equations obtained using an
  independent continuation method~\cite{Ribe2004}.  Arrows denote
  transient regimes observed in the dynamic simulation when the system
  jumps to a different solution branch after encountering a limit
  point (by contrast the continuation method records a steady but
  unstable solution, corresponding the part of the curve with negative
  slope).}
\label{fig:validation}
\end{figure}

The numerical code was validated by comparing its predictions of the
steady coiling frequency to the predictions of the continuation method
of Ribe \cite{Ribe2004}.  The agreement is excellent for all fall
heights (Figure \ref{fig:validation}).  The hysteresis of the dynamic
simulation in the range $1.1\leq \hat H\leq 2.2$ is physical, but
inaccessible to the continuation method because the latter records all
steady-state solutions regardless of their stability.

\section{Simulation results}
\label{dimensionless}

Our dynamic simulations of the sewing machine patterns, based on the
DVT method were carried out with non-dimensional quantities.  This is
achieved by setting the density $\rho$, the viscosity $\mu$, and the
gravity $g$ to the value 1.  This choice implies that both the length
scale $(\nu^2/g)^{1/3}$ and the time scale $(\nu/g^2)^{1/3}$ of the
problem introduced in equation (\ref{hhat}) are equal to 1.  The three
other physical parameters namely the nozzle diameter, the flow rate
and the surface tension were chosen to match the values of
$\Pi_1=670$, $\Pi_2=0.37$ and $\Pi_3=1.84$ prescribed by Morris et
al.\ experiments \cite{Morris:2008p1754}: $d=0.187$, $Q=
22.9\,10^{-6}$ and $\gamma=1.20\,10^{-3}$.  The simulations were initiated
from a vertical thread of uniform radius comprising 172 segments of
equal length, falling from a height $\hat H = 0.86$ (gravitational
regime) onto a belt at rest.  To avoid dealing with a shock when the
thread hits the belt, the simulation was started with the bottom end
of the thread clamped into the ground.  The simulation was run until
the radius settled to a steady profile as a function of the elevation,
and a steady state of coiling was established.  Next the $(\hat H,
\hat V)$ space was sampled by slowly varying $\hat H$ or $\hat V$ in
turn. 

\begin{figure}
\centerline{\includegraphics[%
width=.6\textwidth
]{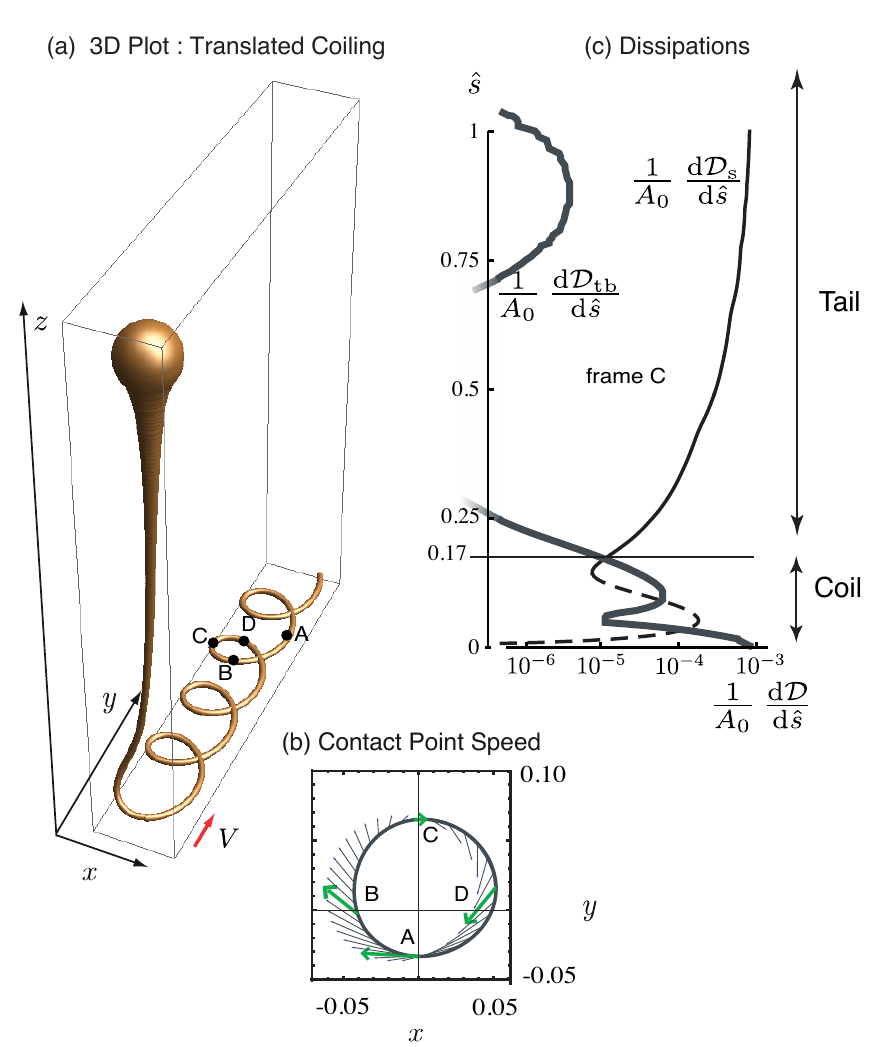}} 
\caption{Simulation of the translated coiling pattern for
$\Pi_1=670$, $\Pi_2=0.37$, $\Pi_3=0$, $\hat{H}=0.98\,$ and
$\hat{V}=0.022$ using 182 vertices for 181 segments.  (a)
Three-dimensional perspective view.  A--D denote reference points
along the thread.  (b) Thick gray curve: trajectory of the contact
point in the frame of the nozzle.  The velocity of the contact point
relative to the belt is shown at various times (thin gray lines) and
highlighted when the contact points passes the reference points A--D
(green arrows) .  (c) Viscous power dissipated per unit length
$\mathrm{d}\mathcal{D}/\mathrm{d}\hat{s}$ by the stretching modes
(plain black and dashed black curves, respectively), and by the
twisting and bending modes (thick grey curve), as a function of
arc-length $\hat{s}$, when the contact point passes the apical
reference position C.
All numerical quantities are dimensionless, as explained in the
beginning of section~\ref{dimensionless}.}
\label{fig:speed}
\end{figure}

As an illustration of the capabilities of our numerical technique, we 
show in Figure~\ref{fig:speed} a simulation of the 'translated coiling' pattern
that occurs for relatively low belt speeds.
Fig.~\ref{fig:speed}a shows a three-dimensional view of the falling
thread and the trace it lays down on the belt.  The simulation time is
$0.73\;\mathrm{s}$ for one period of this pattern using a
$2.6\;\mathrm{Ghz}$ Intel Core i7 processor and $8~\mathrm{Go}$ of 
$1067\;\mathrm{Mhz}$ DDR3 memory.

Fig.~\ref{fig:speed}b shows the trajectory of the contact point in the
frame of the nozzle (solid line).  Interestingly, the belt motion
breaks the symmetry of steady coiling not only in the longitudinal
$x$-direction, but also in the transverse $y$-direction.
Fig.~\ref{fig:speed}b also shows the velocity of the contact point
relative to the moving belt as a function of position along the
trajectory (thin lines and green arrows).  The magnitude of the
relative velocity varies significantly over a period, in contrast to
the meandering pattern for which it is nearly uniform
\cite{Morris:2008p1754}.  The relative speed is maximum at point A
where the contact point is moving upstream along the belt, and very
small at C where the motion is downstream.  Finally,
fig.~\ref{fig:speed}c shows the amount of viscous power dissipated by
the various modes, as a function of arc-length along the thread.  The
curves cross each other at a height $z\approx 0.1$ that corresponds to
the transition from the bending-dominated coil, to the
stretching-dominated tail.  Thanks to adaptivity, the coil is well
resolved and the curves for the viscous power dissipation remain
smooth there, even though they vary rapidly.
%

In addition to the dimensionless parameters $\Pi_1$, $\Pi_2$ and $\Pi_3$ in
equation~(\ref{pi123}) that describe the fluid properties and the
ejection conditions, the patterns  depend on the dimensionless fall height $\hat H$ in
equation~(\ref{hhat}), and the dimensionless belt speed
\begin{equation}
    \hat{V}=V (\nu g)^{-1/3}. 
    \label{vhat}
\end{equation}
Our simulations were carried out by varying $\hat H$ and $\hat V$ for
fixed values of $\Pi_1=610$ and $\Pi_2=0.370$ corresponding to the
experiments of Morris et al.\cite{Morris:2008p1754}.  Some of our
simulations were done with a surface tension parameter matching the
experimental value $\Pi_3 = 1.84$; for reasons of numerical stability, however, most of the simulations used $\Pi_3=0$. 
 
\begin{figure*}
    \centerline{\includegraphics{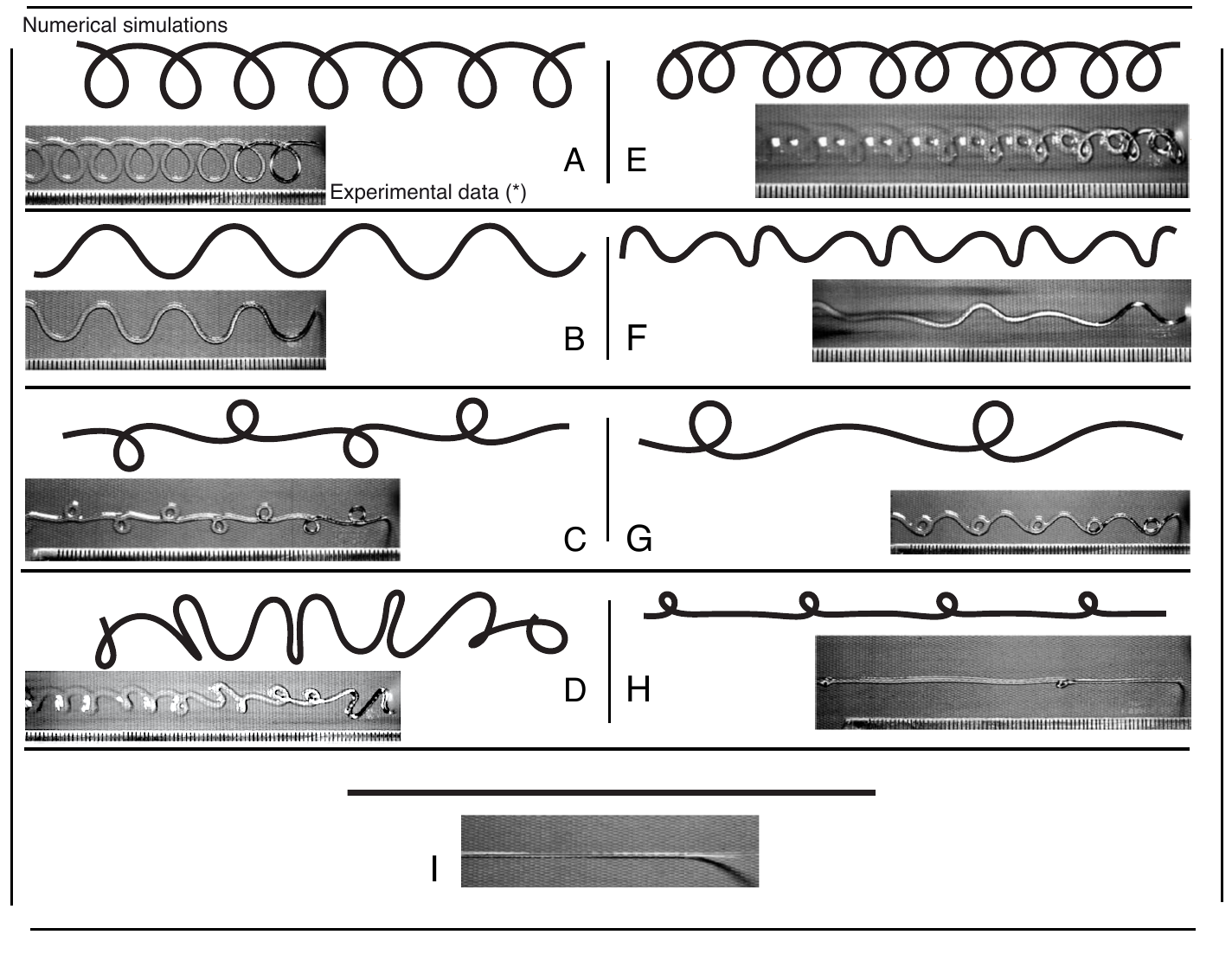}}
    \caption{Simulated sewing machine patterns (black curves) together
    with photographs of similar patterns observed experimentally
    \cite{ChiuWebster:2006p1764}.  The comparison is qualitative
    because the parameters used in the simulations ($\Pi_1 = 670$,
    $\Pi_2 = 0.37$, $\Pi_3 = 0$) are not the same as in the
    experiments of \cite{ChiuWebster:2006p1764}.  A: Transient
    coiling, B: Meanders, C: Alternating loops (figure-of-eight)
    pattern, D: Disorder, E: Double coiling, F: Double meanders, G: W
    pattern, H: Stretched coiling, I: catenary. ($^*$) Photo Courtesy: Chiu-Webster and Lister.}
\label{fig:patterns}
\end{figure*}

Figure~\ref{fig:patterns} summarizes all the types of patterns that
were encountered when scanning the ($\hat{H},\hat{V}$) plane in the
simulations, together with their experimental
equivalents~\cite{ChiuWebster:2006p1764}.
The simulation reproduces nine out of the ten patterns reported by
Morris et al.\cite{Morris:2008p1754} and observed by CWL
\cite{ChiuWebster:2006p1764}.  The only missing pattern, the slanted
loops, will be discussed later on.  To ease the understanding, the
name 'figure-of-8' is replaced by 'alternating loops' which more
accurately describes the pattern in Figure~\ref{fig:patterns}C.

\begin{figure}[!h]
\centerline{\includegraphics[%
width=.6\textwidth
]{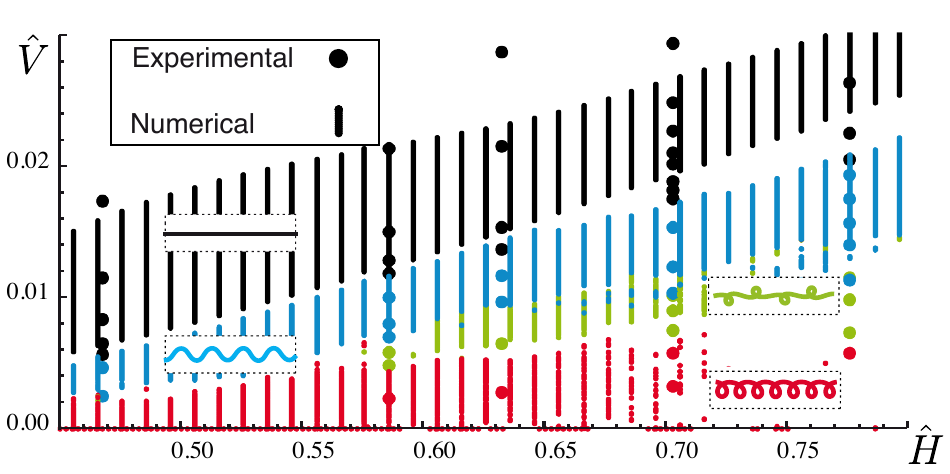}}
\caption{
	Phase diagram of the numerically simulated patterns as a
	function of dimensionless fall height $\hat H\leq 0.80$ and
	dimensionless belt speed $\hat V$, for $\Pi_1 = 670$, $\Pi_2 =
	0.37$ and $\Pi_3 = 1.84$.  The observations of Morris et
	al.~\cite{Morris:2008p1754} are shown by colored dots for
	comparison.  The typical appearance of each pattern (catenary,
	meanders, alternating loops, translated coiling) is shown in insets.}
\label{fig:phased}
\end{figure}

In Figure~\ref{fig:phased} a phase diagram lists all the numerical
patterns encountered in the $(\hat H, \hat V)$ space for $\hat H\leq
0.8$.
The effect of surface tension is included, $\Pi_3 = 1.84$ in
the simulation.  For comparison, the patterns observed experimentally
by Morris et al.\cite{Morris:2008p1754} are shown by dots.  The
agreement between the simulations and the experiments is remarkable:
all four patterns that were observed experimentally in this region of
the parameter space, namely translated coiling, alternating loops,
meanders and catenary, are recovered.  In addition the location of the
boundaries separating the different patterns are reproduced
accurately.

\begin{figure*}
  \centerline{\includegraphics[width=7in]{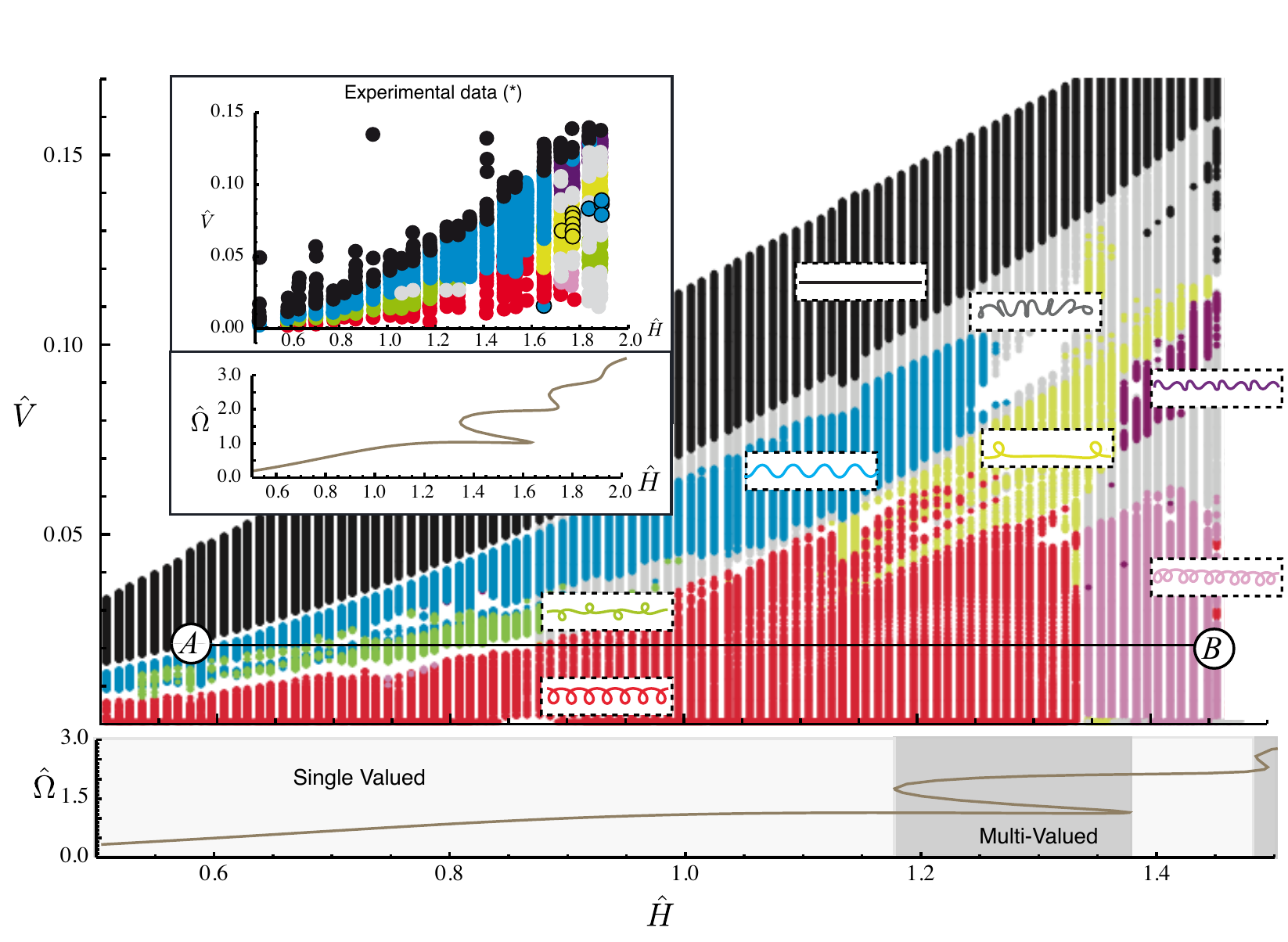}}
  \caption{Phase diagram of sewing machine patterns determined from
  numerical simulations with $\Pi_1 = 670$, $\Pi_2 = 0.37$ and no
  surface tension ($\Pi_3 = 0$).  The patterns simulated include
  translated coiling (red), meanders (blue), alternating loops (green), double coiling (pink), stretched coiling (yellow), double
  meanders (purple), and disordered patterns (grey).  The catenary
  pattern in black would extend all the way up in the diagram, the
  border between black stripes and the white region at the top being
  only the result of the simulation being stopped for large enough
  values of $\hat{V}$. The W-pattern (yellow dots circled in black in inset) and slanted loops
   (blue dots circled in black in inset) are discussed in section~\ref{patterns}.
    The horizontal path AB is used to construct
  Figure~\ref{traceOmega}.  The coiling frequency $\hat\Omega$ for
  steady coiling is shown as a function of height $\hat H$ below the
  phase diagram.  Inset: Phase diagram determined experimentally
  \cite{Morris:2008p1754} for $\Pi_1 = 670$, $\Pi_2 = 0.37$, and $\Pi_3 = 1.84$
  (top) together with the corresponding curve $\hat\Omega(\hat H)$
  (bottom). }
\label{fig:phasedF}
\end{figure*}

The inclusion of surface tension gives rise to numerical instabilities
for heights approximately above $\hat H > 0.70$, which we have not
been able to overcome by decreasing the mesh size or the time-step.
This is why there is no simulation data shown in the lower right
portion of Figure~\ref{fig:phased}.  Since surface tension is not
expected to modify qualitatively the dynamics of threads (see
Fig.~\ref{fig:steady}) we investigated the case of larger fall heights
without any surface tension ($\Pi_3= 0$).  Five new patterns were
observed for larger fall heights, as shown in Figure
\ref{fig:phasedF}, namely double coiling, double meanders, stretched
coiling, W-pattern, disordered pattern, despite the absence of surface
tension in the simulations.  The new portion of the phase diagram,
$\hat H > 0.8$, is very similar to that reported by Morris et
al.\cite{Morris:2008p1754}, shown in the inset in
Fig.~\ref{fig:phasedF}.  In both diagrams, the alternating loops
pattern disappears at a critical height, beyond which there is a
substantial height `window' containing only simple patterns (catenary,
translated coiling, and meanders).  When the height is increased,
three patterns having a complex aspect (disordered pattern, stretched
coiling and the double meanders) all appear together at nearly the
same height.  Finally, for some values of the height, disordered
patterns, shown in grey in the diagram, occur in two separate ranges of
the belt speed, with stretched coiling in between.

It is instructive to compare the numerical and experimental phase
diagrams with the curves $\hat\Omega_c(\hat H)$ of frequency vs.\ height
for steady coiling, calculated with same value of surface tension
($\Pi_3 = 0$ for the simulations, $\Pi_3 = 1.84$ for the experiments).
These curves are shown below each phase diagram in
Figure~\ref{fig:phasedF}.  The comparison reveals that some of the
more complicated patterns (stretched coiling, W-pattern, disordered
pattern) appear at heights close to that for the onset of the
multivalued IG regime in the steady coiling.  In the steady
coiling geometry, it is known that the multivalued regime is caused by
the competition of several `viscous pendulum' modes.  This
suggests that the complex patterns of the sewing machine are produced
by the non-linear interaction of these pendulum modes.

Despite their similarities, the numerical (N) and experimental (E)
phase diagrams in Figure~\ref{fig:phasedF} exhibit some systematic
differences.  In E, double coiling (pink) first appears at the same
height as the disordered (grey) and stretched coiling (yellow),
whereas in N it appears at significantly greater heights.  Double
meanders (purple) have a common boundary with the catenary pattern
(black) in E, but occur only for significantly lower belt speeds in N.
In N, the catenary state can transition to disorder (grey) over a
range of heights, unlike in E. The range of belt speeds for double
coiling is significantly wider in N than in E. Finally, in N the
W-pattern is observed sporadically and for greater heights than in the
diagram in Fig.~\ref{fig:phasedF}.  Some of these differences are due
to the absence of surface tension in the simulations, and to the fact
that collisions of the viscous thread with its trace are not
prevented.  Another explanation for the discrepancies may be the fact
that Morris et al.~\cite{Morris:2008p1754} performed their pattern
recognition visually, whereas we used a more quantitative automatic
procedure based on Fourier decomposition.  This is described in the
following section, where we review each of the individual patterns in
detail and propose a systematic classification of them.

\section{Analysis of the patterns}

\begin{figure}
  \centerline{\includegraphics[%
width=.48\textwidth
]{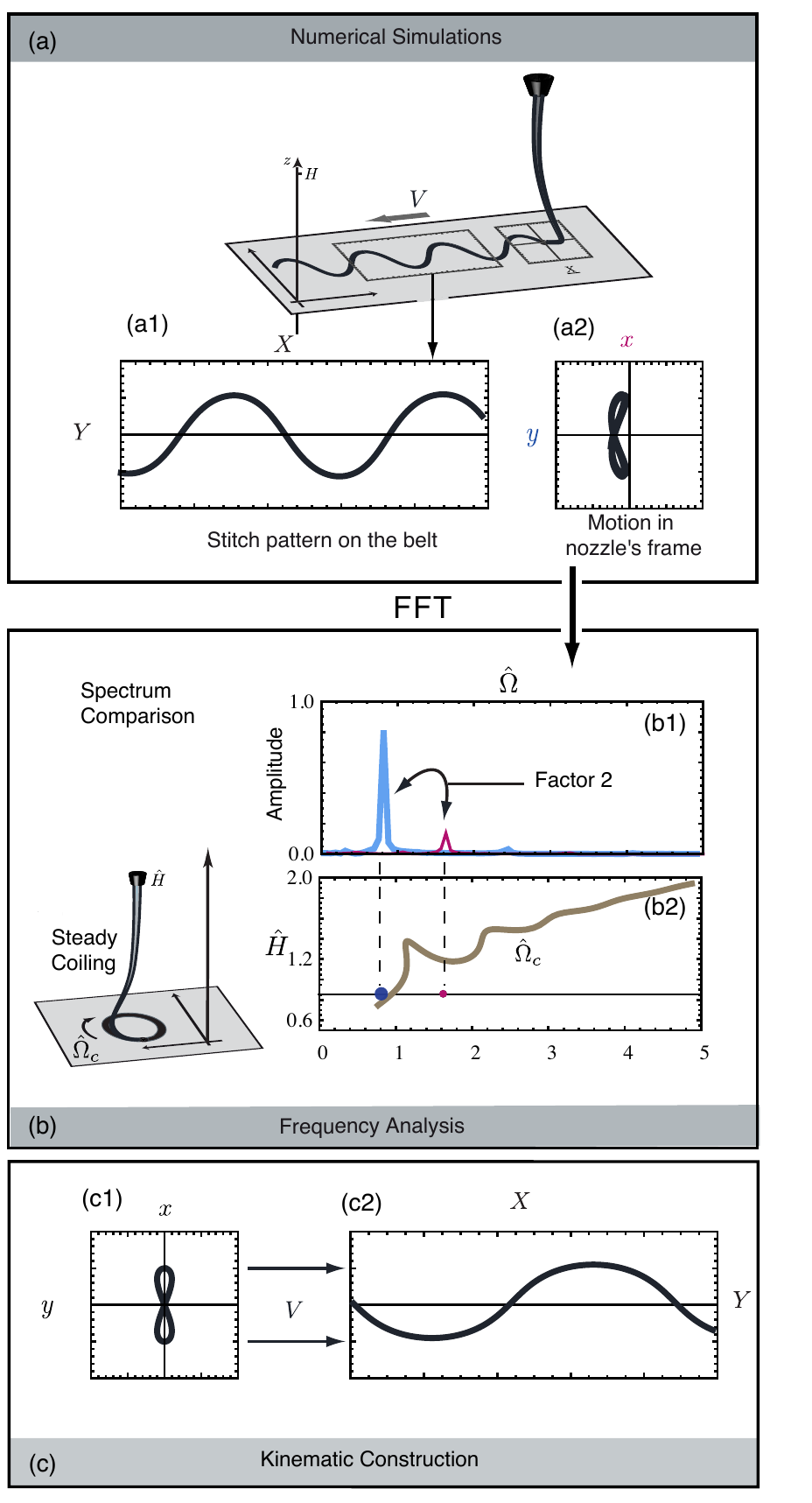}}
  \caption{Three-stage analysis of the patterns, illustrated for the
  case of meandering.  (a) From the trace of the thread on the belt
  (a1), we extract the trajectory of the contact point in the frame of
  the nozzle (a2).  (b1) Fourier spectra of the longitudinal (red)
  and transverse (blue) components are obtained by Fast Fourier
  Transform, and compared to the frequencies $\hat\Omega$
  of steady coiling for the same
  height: the blue and red dots in (b2) are the main frequencies in
  the simulation, while the intersection of the horizontal line with
  the thick black curve is the frequency of steady coiling.  In
  the example shown here the belt speed is close to the critical
  value for the onset of meandering and the dominant transverse frequency  $\omega_1^{(y)}$ is close to the steady coiling frequency $\hat{\Omega}_c$
  for the same height.  (c) The two main frequencies extracted by FFT are
  injected into the kinematic model (\ref{kinmodel}) to generate a
  synthetic motion for the contact point in the nozzle frame (c1), and
  a synthetic stitch pattern (c2).  In the example shown, the motion
  involves only one transverse frequency $\omega_1^{(y)}$ and one
  longitudinal frequency $\omega_1^{(x)} = 2\omega_1^{(y)}$.}
\label{ExempleFFT}
\end{figure}

To illustrate our pattern analysis, we fix the belt speed
$\hat V = 0.02$ and vary the fall height, thereby following the
horizontal line AB in the phase diagram of Fig.~\ref{fig:phasedF}.
In order of increasing height, the patterns seen along this line are
meanders, alternating loops, translated coiling, and double coiling.
We track the spectral content of these patterns continuously as they
change smoothly or bifurcate.
To do so, we focus on the trajectory of the contact point of the
thread with the belt.  Let $x(t)$ and $y(t)$ be its longitudinal and
transverse coordinates in the laboratory (nozzle) reference frame, and
define $\mathbf{x}(t)=x(t)+iy(t)$.  Let $X(t, t^*)$ and $Y(t, t^*)$ be
the coordinates (also relative to the nozzle) at time $t$ of a
material point that was laid down on the belt at time $t^*<t$, and
let $\ve X(t,t^*)=X(t,t^*)+i\,Y(t,t^*)$ be a generic point in the
trace.  The advection by the belt is expressed by:
\begin{equation}
\ve X(t,t^\star)=
\mathbf{x}(t^\star)+ (t - t^*)\,V.
\label{bigx}
\end{equation}
This equation means that the pattern $\ve X(t,t^*)$ seen on the
belt is obtained by unfolding the motion of the contact point motion
$\ve x(t)$, as illustrated in Fig.~\ref{ExempleFFT}a.

The numerical traversal of the line AB in Figure \ref{fig:phasedF}
required about 78500 time steps of size $\delta t=0.1(\nu/g^2)^{1/3}$.
We performed a Fourier analysis of the motion $\ve x(t)$ over a
sliding window of 2000 time steps (the FFT spectrum was computed every
500 steps).  The spectra obtained in this way typically comprise
several well-defined peaks whose frequencies can be determined
precisely (Fig.~\ref{ExempleFFT}b).

 \begin{figure}[!h]
  \centerline{\includegraphics{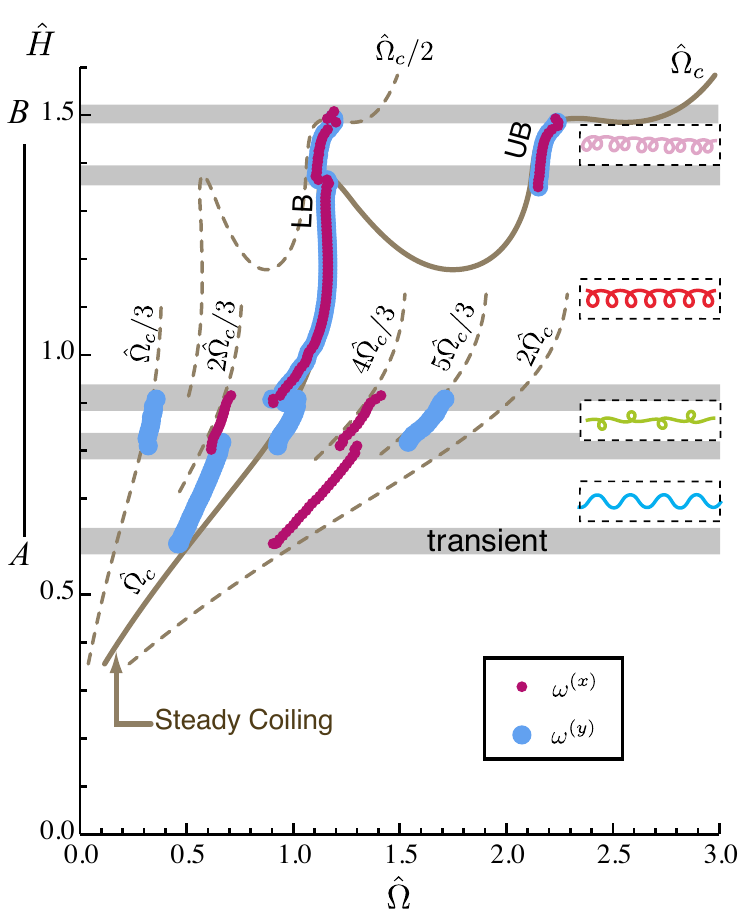}}
\caption{Frequency content of patterns encountered along the line AB
through the phase diagram of Figure \ref{fig:phasedF}.  Frequencies of
transverse ($y$-direction) and longitudinal ($x$-direction) motion are
shown as blue and red, respectively.  Note that these frequency are the same
in the upper region of the diagram where the two colors overlap.  In
the dashed insets, the patterns are identified using the same color
codes as in Figure~\ref{fig:phasedF}.  Grey bands indicate ranges of
heights for which the patterns are transient.  Also shown is the
frequency $\hat\Omega_c$ of steady coiling as a function of height
(brown solid line), together with several multiples of that frequency
(brown dotted lines).}
\label{traceOmega}
\end{figure}

Let $\omega^{(x)}_1$, $\omega^{(x)}_2$, etc.\ be the peak frequencies
of the motion in the $x$-direction, and $\omega^{(y)}_1$,
$\omega^{(y)}_2$, etc.\ be those for the motion in the $y$-direction.
Because the fall height $\hat H$ is slowly changing with time during
the simulation, each observed frequency $\omega^{(x)}_i$ or
$\omega^{(y)}_i$ can be plotted on a diagram
to provide a `portrait' of the evolving frequency content of the
contact point motion, as a function of the fall height $\hat H$.
The result is shown in Figure \ref{traceOmega}.  The principal
observed frequencies $\omega^{(x)}_n$ and $\omega^{(y)}_n$ are
indicated in red and blue, respectively.  Also shown for reference is
the steady coiling frequency $\hat\Omega_{c}(\hat H)$ for the same fluid
properties and ejection parameters (solid line), together with several
multiples (1/3, 2/3, 1/2, 4/3, 5/3, 2) of that frequency (dashed
lines).

The first pattern ($0.62 \leq\hat H\leq 0.8$) is meandering, which is
characterized by two frequencies with a ratio
$\omega^{(x)}_1/\omega^{(y)}_1 = 2$.  At the lowest height $\hat H =
0.62$ where meandering first appears, the meandering frequency
$\omega^{(x)}_1$ is very close to the steady coiling frequency
$\hat{\Omega}_c$ for the same height, as expected on theoretical grounds\cite{Ribe:2006p691, blount11}.  The next pattern ($0.8 \leq\hat
H\leq 0.9$) is the alternating loops, which has a rich spectrum
involving five multiples of $\hat\Omega_c/3$.  Translated coiling appears
next ($0.9 \leq\hat H\leq 1.35$), and is characterized by a single
frequency $\omega^{(x)}_1 = \omega^{(y)}_1$ very close to the steady
coiling frequency.  Finally, double coiling ($1.35 \leq\hat H\leq
1.5$) has two frequencies $\omega^{(x)}_1 = \omega^{(y)}_1\approx
\hat\Omega_c$ and $\omega^{(x)}_2 = \omega^{(y)}_2\approx \hat\Omega_c/2$.

Figure~\ref{traceOmega} shows that the stitch patterns are
combinations of motions in two orthogonal directions with frequencies
closely related to the steady coiling frequency $\hat\Omega_c$.
Accordingly, we now change our perspective and classify the patterns
based on their frequency content rather than on the shape they lay
down on the belt.  This frequency analysis is used to set up an
efficient tool for identifying the patterns and assembling the
numerical phase diagram automatically.  In addition it leads to a simple
kinematic model that provides a unified description of the different
patterns.

\subsection{Spectral signature of a pattern}

We illustrate our method using the example of meandering which in most
cases is the first pattern to bifurcate from the catenary state as the
belt speed decreases.  The red and blue lines in
figure~\ref{ExempleFFT}b show typical spectra of the motion of the
contact point in the longitudinal $x$ and transverse $y$-directions,
respectively.  The amplitude of the transverse motion is much greater
than that of the longitudinal motion, and the frequency of the latter
is exactly twice that of the former, $\omega_1^{(x)} =
2\omega_1^{(y)}$.  This suggests that a meander pattern can be
synthesized by retaining only the two main frequencies, viz.\
\begin{equation}
\ve x(t) = \alpha_1\cos(2 \omega_1^{(y)} t) + i \,  \beta_1 \cos(\omega_1^{(y)}+\pi/4) 
\label{twofreq}
\end{equation}
where $\alpha_1$ and $\beta_1$ are the amplitudes of the longitudinal
and transverse motions, respectively.  Here the phase difference $\pi/4$ is
required to reproduce the symmetry of the pattern.  A similar
two-frequency model was used by Morris et al.~\cite{Morris:2008p1754}
to analyze weakly non-linear meanders.
Figure~\ref{ExempleFFT}c1 shows the contact line trajectory in the
frame of the nozzle implied by (\ref{twofreq}) with
$\alpha_1/\beta_1=0.2$, and Figure~\ref{ExempleFFT}c2 shows the
corresponding meander pattern obtained by advecting the motion
(\ref{twofreq}) in the $x$-direction with $V=1.4(\nu g)^{1/3}$ and
$\omega_1^{(y)}=1$.
Based on Figure~\ref{ExempleFFT}, we define as a `meander' any pattern
whose longitudinal motion, compared to the transverse motion, has 
twice the frequency, a much smaller amplitude, and a phase shift of 
$\pi/4$.

Generalizing the above example, we show that all the sewing machine
patterns can be represented by a superposition of a few harmonic
motions of the form:
%
\begin{equation}
    x(t)+i y(t)  =  \sum_{j=1}^{N_x}\alpha_j \cos(\omega^x_{j}
    t+\phi^x_j) + i \sum_{j=1}^{N_y} \beta_j \cos(\omega^y_{j} 
    t+\phi^y_j)
    \textrm{.}
\label{kinmodel}
\end{equation}
Two terms in each direction are sufficient to reproduce the main
features of the patterns, $N_x\leq 2$ and $N_y\leq 2$.  In
Eqn.~(\ref{kinmodel}), $\alpha_j$ and $\beta_j$ are the amplitudes of
the components of the motion with frequencies $\omega^{(x)}_j$ and
$\omega^{(y)}_j$, and $\phi^x_j$ and $\phi^y_j$ are the phases
relative to the highest-frequency mode.  We now show how each of the
sewing machine patterns can be characterized in terms of the
parameters that appear in Eqn.~(\ref{kinmodel}).

\subsection{Identification of the patterns}
\label{patterns}

The identity of each pattern is determined not by the absolute values
of the parameters in Eqn.~(\ref{kinmodel}), but rather by the
dimensionless groups that can be formed from them, namely frequency
ratios, amplitude ratios, and the relative phases $\phi_j$.  In the
following we identify the characteristic values of these groups for
each of the patterns in turn.  For ease of reference, the results are
summarized in Table \ref{table:kinematic}. \\

\begin{figure*}[!h]
  \centerline{\includegraphics[width=14cm]{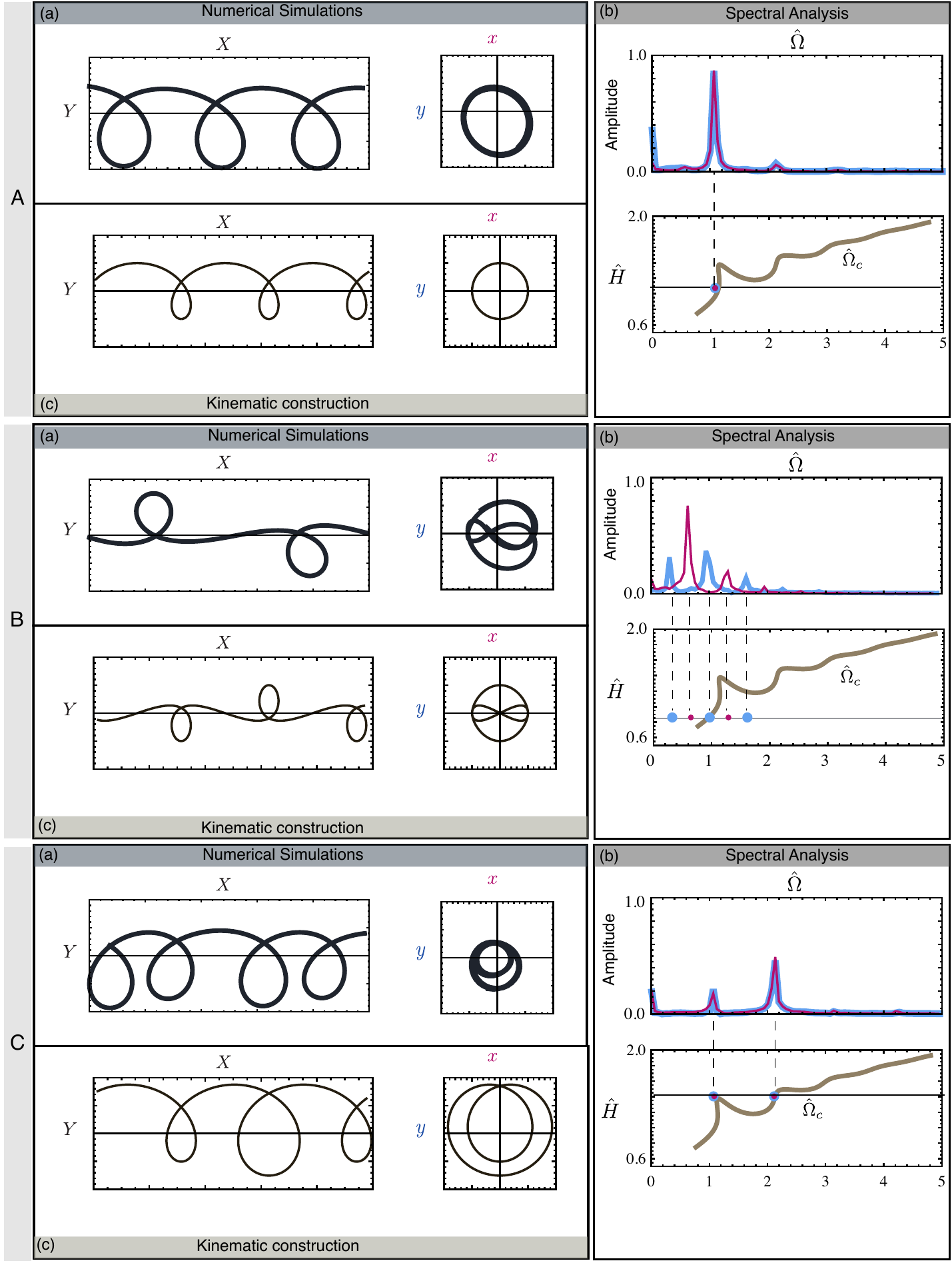}}
  \caption{Kinematic analysis of individual sewing machine patterns.
(A) coiling; (B) alternating loops; (C) double coiling. For each pattern,
the upper left, right, and lower left panels correspond to
parts a), b) and c), respectively, of Figure~\ref{ExempleFFT}.
The kinematical reconstruction is a proof of concept, there is no attempt to match the wavelength of the simulations. 
}
\label{fig:final}
\end{figure*}

1.  \textit{Translated coiling}.  This pattern occurs for $0.5
\leqslant \hat H \leqslant 1.35$ and low belt speeds
(Figure~\ref{fig:phasedF}).  Figure~\ref{fig:final}A shows a
simulation of this pattern (upper left) and the corresponding Fourier
spectra of the longitudinal and transverse motions of the contact
point (upper right), and the steady coiling frequency $\hat\Omega_c$.
The longitudinal and transverse motions have similar amplitudes and
are characterized by a single dominant frequency $\omega_1^{(x)} =
\omega_1^{(y)}$ that is very close to the steady coiling frequency
$\hat\Omega_c(\hat{H})$ for the same fall height $\hat{H}$.  The peak
frequency deviates from its original value as the belt speed
increases.  The amplitudes in both directions remain equal and an
almost circular shape is created.  The panels at lower left show the
reconstructed motion in the frame of the nozzle (right) and on the
belt (left), calculated using (\ref{kinmodel}) with $N_x = N_y = 1$,
$\alpha_1 =\beta_1 = 1$ and $\omega_1^{(x)} = \omega_1^{(y)} = 1$.
Note that the experimental pattern shifts upwards or downwards as the
belt speed is increased; this shift does not affect the spectral
context, but could be taken into account by including a purely
imaginary constant in Eqn.~(\ref{kinmodel}).
%
\\

\begin{table*}
    \centering  
    \begin{tabular}{|l||cc|cc|cc|cc||cc|cc|c|ccc} 
	\hline
	Patterns & $\omega_1^x$ & $\omega_2^x$ & $\omega_1^y$  &
	$\omega_2^y$ & $\phi_1^x$ &  $\phi_2^x$&$ \phi_1^y$&  $\phi_2^y$& 
	$\alpha_1$&  $\alpha_2$& $\beta_1$& $\beta_2$&  V\\ 
	\hline
	Translated Coiling & $1^{*} $& - & $1^{*}$ & - &0&- &0&-&1.&-&1.&-&.5  \\ 
	Meanders & 2 & - & $1^{*} $& - & 0 & - &$\pi/4$ & -
	&  .2 & - &1. & - & 1.4  \\
	Alternating loops &1 & - & 1/2 & $3/2^* $ &$\pi/2$ & - &0 &0 & 1 & - &.5&.5&0.33\\
	Double Coiling	& 1/2 &$1^*$ & 1/2 & $1^*$& $\pi/2$ &$\pi/2$ & 0 & 0 & .5 &1.5&.1&1.5&.5\\
	Double Meanders &1/2&-&-&$1^* $& $\pi/4$ & - & 0 & - &1.&0&0&1.5& .75\\
	Stretched Coiling & 1 & $2^*$ & 1 & $2^* $ & $\pi/2$ &$\pi/2$&0&0 &1. & .1 & .5 & .1  & .6 \\
	W-pattern & 1 & $2^*$ & 1 & $2^* $ &  $\pi/2$ &$\pi/2$&0&0 & 1.& .2 & .2 & .5 & .7\\
	Catenary & - & - & 0 & - & - &- & 0 & - & - & - & 0 & - & 1 \\
	\sout{Disorder}& -  & - &- &  -&  -& -& -& - & - & -& - & -& - \\
	\hline
    \end{tabular}
    \caption{Parameters of the kinematic model in
    Eqn.~(\ref{kinmodel}) used to construct synthetic patterns.
    Because the patterns are defined by the relative values of the
    frequencies $\omega_1^x$ and $\omega_1^y$, the frequency 1 is
    assigned by convention to the peak of largest amplitude.  The
    other frequencies are given by ratios of small integer numbers.
    A star indicates a frequency that is locked to the steady coiling
    frequency $\hat\Omega_c$.  Likewise, the amplitudes $\alpha_{1}$,
    $\alpha_{2}$, $\beta_{1}$ and $\beta_{2}$ are given relative to
    each other, and correspond to typical values.  $V$ indicates the
    speed of the belt used to unfold the synthetic patterns.  A dash
    indicates that the parameter is not relevant for the pattern in
    question.  The disordered pattern is not reconstructed using the
    kinematical model as it involves more harmonics.}
  \label{table:kinematic}
\end{table*}

2.  \textit{Meanders}.  On Figure~\ref{fig:phasedF} this pattern is
seen for $ 0.6 \leqslant \hat H \leqslant1.3 $ and a range of
intermediate belt speeds.  The typical Fourier spectra of the
meandering pattern were previously shown in Figure~\ref{ExempleFFT}.
The pattern is a superposition of a single longitudinal, and a single
transverse harmonic motions, with frequency ratio
$\omega_1^{(x)}/\omega_1^{(y)} = 2$, amplitude ratio
$\alpha_1/\beta_1\ll 1$, and a relative phase $\phi_1^y = \pi/4$ (with
the convention $\phi_{1}^x=0$).  Near the catenary/meander boundary
$\omega_1^{(y)} \simeq \hat\Omega_c$; farther from the boundary, the
meandering frequency departs significantly from $\hat\Omega_c$.  The
regular symmetrical meanders correspond to a phase difference of
$\pi/4$ between the two directions (fig.~\ref{ExempleFFT}) However,
the pattern may deform into a bean-like shape in certain cases.  This
situation was reproduced kinematically by reducing the phase difference
to a value close to $\pi/6$. \\

3.  \textit{Alternating loops}.  This pattern was called
`figure-of-eight' by CWL\cite{ChiuWebster:2006p1764} and Morris et
al.\cite{Morris:2008p1754}, but we prefer to call it `alternating
loops'.  The domain of this pattern is an elongated `bubble'
sandwiched between translated coiling and meandering at relatively low
fall heights $\hat H\leq 1$ (Figure~\ref{fig:phasedF}).  This pattern
displays a remarkably rich frequency spectrum with five principal
peaks (Figure~\ref{fig:final}B-c).  In contrast to meandering, the
motion with the largest amplitude is longitudinal, with a frequency
$\omega_1^{(x)}$ that locks onto the frequency $2\hat\Omega_c/3$ (see
also Figure~\ref{traceOmega}).  The next largest peaks correspond to a
transverse motion with frequencies $\omega^{(y)}_2=\hat\Omega_c$ and
$\omega^{(y)}_1=\hat\Omega_c/3$, both amplitudes being very close.
The harmonics $4\hat{\Omega}_{c}/3$ in the longitudinal direction, and
$5\hat{\Omega}_{c}/3$ in the transverse direction are also visible.


The frequencies of all five peaks can be written compactly as
\begin{alignat}{2}
\omega^{(y)}_p& = (2p-1)\,\frac{\hat\Omega_c}{3} & \quad  & (p=1,2,3),
\nonumber
\\
\omega^{(x)}_p & = 2p\,\frac{\hat\Omega_c}{3} & &  (p =1,2). 
\nonumber
\end{alignat}
Even though the spectra in Fig.~\ref{fig:final}B-b shows five peaks,
the kinematic model in Eqn.~(\ref{kinmodel}) generates an almost
identical pattern if one retains only the three main contributions
$\omega_1^{(x)}$, $\omega_1^{(y)}$, and $\omega_2^{(y)}$ with an
amplitude ratio $\beta_1/\alpha_1\approx .5$, $\beta_1/\alpha_2\approx
.5$, and phases $\phi_1^x = \pi/2$ and $\phi_1^y = \phi_2^y = 0$
(Figure~\ref{fig:final}B-c).  We used these characteristics of the
three main frequencies as a criterion for automatic detection of
alternating loops.  \\

\begin{figure*}[!h]
  \centerline{\includegraphics{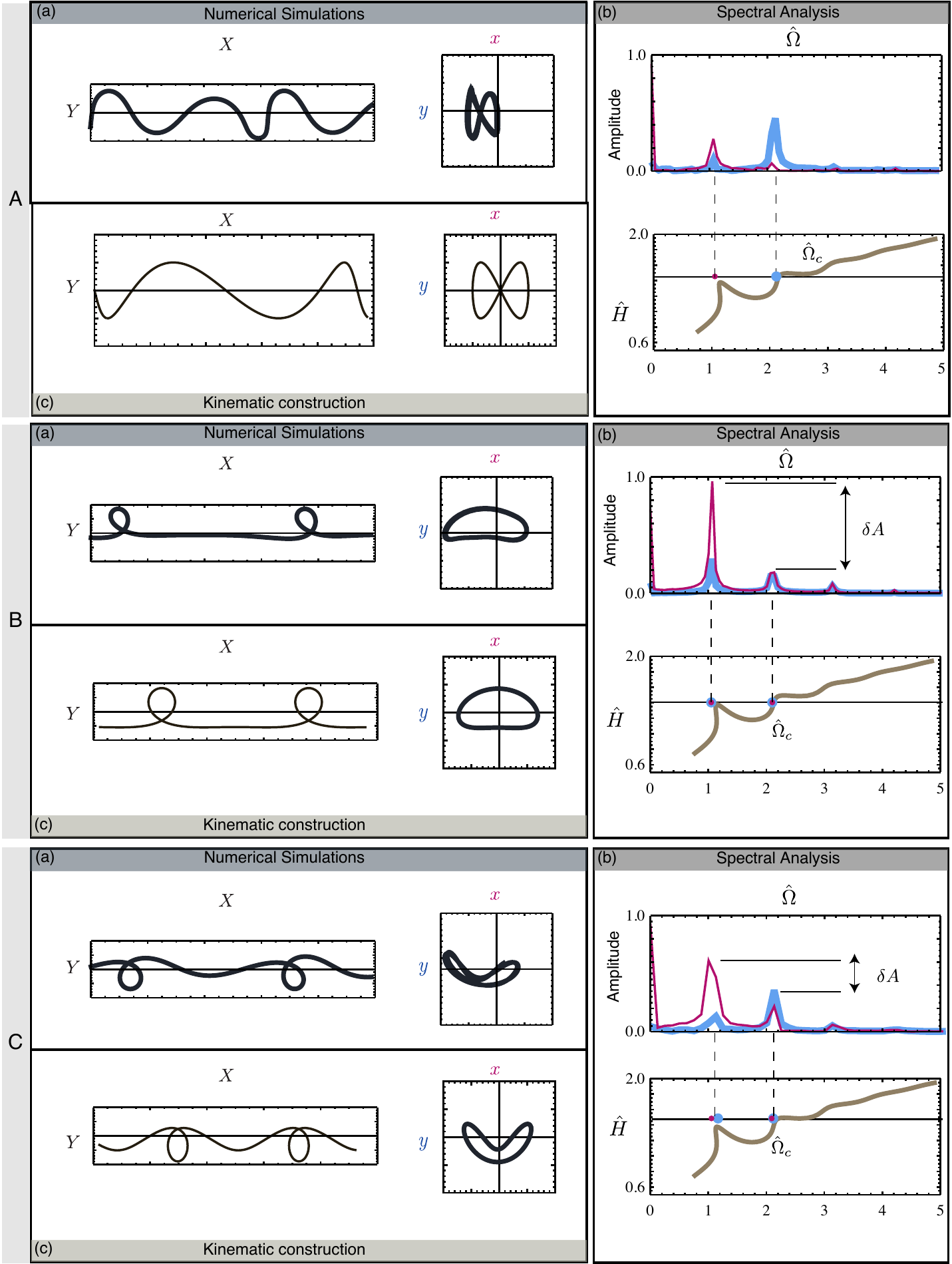}}
  \caption{Same as Figure~\ref{fig:final}, but for (A) double meanders, 
  (B) stretched coiling, and (C) the 'W-pattern'. $\delta A$ denotes the difference of amplitude.}
\label{fig:final2}
\end{figure*}

4., 5.  \textit{Double coiling and double meanders}.  The Fourier
spectra for these patterns are shown in Figures~\ref{fig:final}C and
\ref{fig:final2}A, respectively.  Both the longitudinal and transverse
components have peaks at two frequencies $\omega_1^{(x)} =
\omega_1^{(y)} =\hat \Omega_c/2$ and $\omega_2^{(x)} = \omega_2^{(y)}
= \hat\Omega_c$.  The origin of these frequencies is clear from the
uppermost part of Figure~\ref{traceOmega}, which shows them as
functions of fall height for double coiling.  The range of fall
heights in question is within the inertio-gravitational regime of
steady coiling, for which the curve $\hat\Omega_c(\hat H)$ is
multivalued in specific height ranges (Figure~\ref{fig:steady}).  The portion
$1.2\leq \hat H\leq 1.5$ of that curve has two stable branches
corresponding to different `pendulum' modes of the tail: a lower
branch (labelled LB in Figure~\ref{traceOmega}) with
$\hat\Omega_c\approx 1.15$, and an upper branch (UB) with
$\hat\Omega_c\approx$ 2.1-2.2.  Figure~\ref{traceOmega} shows that the
higher double coiling frequency $\omega_2^{(x)} = \omega_2^{(y)}$
stays locked to the upper branch of $\hat\Omega_c(\hat H)$,
which is the only stable one when $\hat H\geq 1.37$.  The lower
frequency, by contrast, follows a `phantom' branch with frequency
$\hat\Omega_c/2$ that is very nearly a direct continuation of the
lower branch of $\hat\Omega_c(\hat H)$ to greater fall heights.  This
behavior is possible because the ratio of the frequencies of the upper
and lower branches happens to be quite close to 2.0.
 
Although double coiling and double meanders have the same frequency
content, they are distinguished by the relative amplitudes and phases of
the transverse and longitudinal motions. For double coiling, the amplitudes
of the two motions are the same at both frequencies
($\alpha_1=\beta_1$, $\alpha_2=\beta_2$), and 
the relative phases are $\phi_1^x=\phi_2^x=\pi/2$ and $\phi_1^y=\phi_2^y=0$.
For double meandering, by contrast, the transverse motion is
dominated by the frequency $\omega_2^{(y)} = \hat\Omega_c$
while the longitudinal motion is dominated by  
$\omega_1^{(x)} = \hat\Omega_c/2$. The relative phases are 
$\phi_1^x=\pi/4$ while $\phi_1^y=0$.  \\

6., 7.  \textit{Stretched coiling and the W-pattern}.  These patterns
occur predominantly in the range of heights corresponding to
inertio-gravitational coiling (right-hand side of Figure
\ref{fig:phasedF}).  Their typical Fourier spectra are shown in Figure
\ref{fig:final2}B and C, respectively.  Like double coiling and double
meanders, their characteristic signature is the ($\hat\Omega_c/2,
\hat\Omega_c$) frequency couple.  But whereas double coiling and
double meanders are dominated by transverse motion at the frequency
$\hat\Omega_c$, stretched coiling and the W-pattern are
dominated by longitudinal motion at the frequency $\hat\Omega_c/2$.
The difference between stretched coiling and the W-pattern is only due to different relative amplitudes of the transverse and longitudinal motions denoted $\delta_A$ in Figure~\ref{fig:final2}B-b and C-b.  The $x-$direction is dominant in both cases but the peak at $\hat\Omega_c$  in the $y$-direction  is much closer for the W-pattern than for the stretched coiling. Theses differences cause  the change of invagination between the two patterns (Figure~\ref{fig:final2}B-a and C-a).  
The phase difference between the longitudinal and transverse motions is $\pi/2$ in both cases.  \\

\begin{figure}
    \begin{center}
	\includegraphics{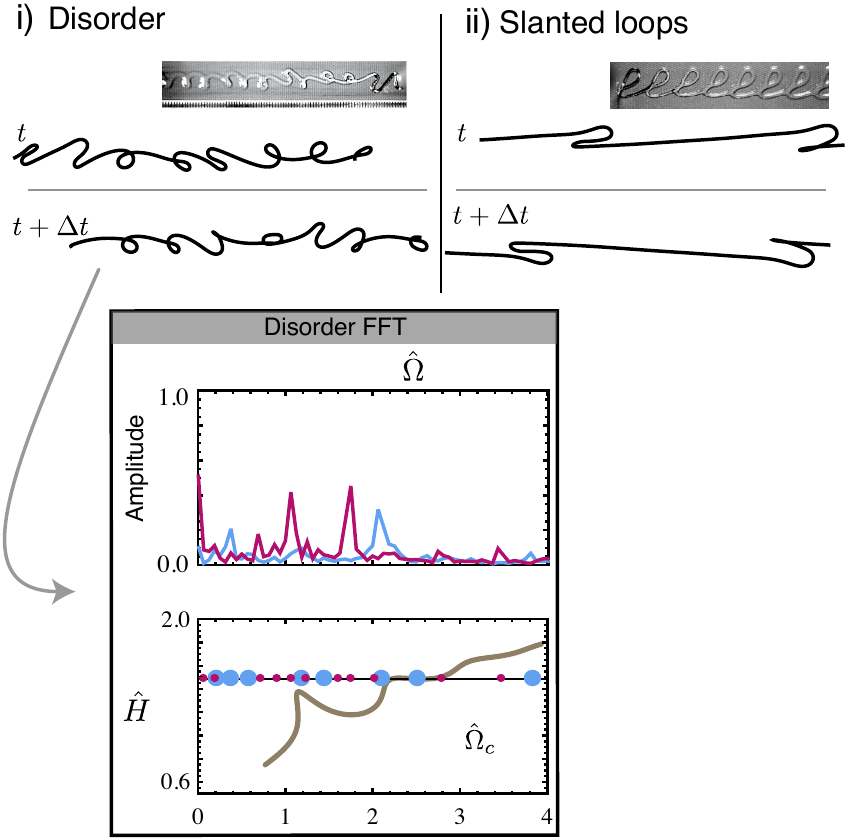}
	\caption{Unsteady patterns: (i) disordered pattern and (ii)
	slanted loops.  Both patterns are shown at two different times
	of the simulations, and compared in the insets to
	experiments~\cite{ChiuWebster:2006p1764}.  The lower part of
	the figure illustrates the FFT of the disordered patterns.}
	\label{disorder}
\end{center}
\end{figure}

8.  \textit{Disorder}.  Disordered patterns appear in several parts of
the phase diagram (grey in Figure~\ref{fig:phasedF}), primarily at
heights within the inertio-gravitational coiling regime.  The typical
Fourier spectra of these patterns is very rich, with more than four
peaks in both the longitudinal and transverse directions with
comparable and strongly time-dependent amplitudes
(Fig.~\ref{disorder}-i and corresponding FFT).  Such
patterns are not transient between two steady patterns, as the
aperiodic behavior persists indefinitely in time.  \\

9.  \textit{Catenary}.  The catenary is obtained when the point of
contact is at rest in the nozzle frame, which happens in the upper
region of the phase diagram.  The FFT spectrum is then empty. \\

In addition to the patterns discussed above, slanted
loops~\cite{ChiuWebster:2006p1764} were also reported by Morris et
al.~\cite{Morris:2008p1754} in a very narrow region of their phase
diagram.  Slanted loops is a pattern wherein a buckle is periodically
laid down on the belt, and subsequently closes up into a loop when the
thread touches itself and coalesces.  Our numerical simulations do not
account for self-contact of the thread, nor for surface
tension-mediated coalescence.  This probably explains why we observed
slanted loops transiently only, as shown in Fig.~\ref{disorder}-ii.  A
similar argument may also explain why the W-pattern is observed for
significantly larger fall heights in the simulations than in the
experiments: considering self-contact of the thread would certainly
favor its existence over the stretched coiling pattern in the
simulation.

We observe that CWL~\cite{ChiuWebster:2006p1764}
reported yet another pattern, side kicks.  Side kicks consist of small
heaps of fluid regularly spaced along an otherwise perfectly straight
trace.  We suggest that this pattern is a limiting case of stretched
coiling in which the amplitude of the transverse motion becomes very
small relative to that of the longitudinal motion.  Side kicks have
not been reported in the experimental phase diagram of Morris and
al., the one we attempt to reproduce in the simulations;
consistently, we have observed this pattern in our
simulation only transiently.

\section{Discussion} 

The simulations of the fluid-mechanical sewing machine presented here
were performed using a new numerical algorithm, Discrete Viscous
Threads (DVT).  The essential idea of DVT is to start from a complete
geometrical and kinematical description of the thread in the discrete
setting, and push the discrete approach as far as possible; in
particular a discrete representation of viscous stress is built based
on a variational view (Rayleigh potentials).  This approach leads to a
code that is robust even for quite large mesh size.  This numerical
method is used to simulate complex unsteady behavior of a viscous
thread and offers good efficiency.  As an example, the traversal
(78500 time steps) of the line AB in Figure~\ref{fig:phasedF} required
$987\;\mathrm{s}$ on a $2.87\,\mathrm{GHz}$ Intel Core processor.  The
accuracy of the method is demonstrated by its ability to reproduce the
curve of steady coiling frequency vs.\ height, as predicted by an
independent continuation method (Figure~\ref{fig:steady}), and by the
close agreement between the calculated and experimentally determined
phase diagrams for the sewing machine patterns
(Figure~\ref{fig:phased}).

At each time step in an unsteady DVT simulation, any desired
kinematical or dynamical variable can be calculated as a function of
arc-length, i.~e.\ at every vertex along the thread's centerline.
Examination of these functions provides insights into the thread's
dynamics.  We saw an example in Figure~\ref{fig:speed}c, which showed
the local rates of viscous dissipation of energy due to deformation by
stretching (or compression), bending, and twisting.  The figure
reveals that the thread is divided into two distinct parts: a `tail'
in which the dissipation is dominated by stretching, and a `coil' in
which it is dominated by bending and twisting, but with a significant
contribution from compression.  The differential equations describing
bending are of higher order than those describing stretching, so
Figure~\ref{fig:speed}c implies that the coil is an `inner' solution or
boundary layer whose presence is required by the need to satisfy all
the relevant boundary conditions at the thread's contact point with
the belt.  While the boundary-layer character of the coil region has
long been recognized for steady coiling \cite{Mahadevan:1998p1053,
Ribe2004}, our simulations open up the possibility of studying the
associated non-steady dynamics.

Another benefit of the simulations is to allow exploration of regions
of parameters space that are inaccessible in the real world but
provide new insights into the dynamics of the thread.  In
Figure~\ref{fig:speed}c, the rates of viscous dissipation for the
bending and twisting modes were added together.  A real thread is made
up of an incompressible fluid, and the ratio of the bending to the
twisting modulus is always $3/2$ (both moduli are in particular
proportional to the fourth power of the thread's radius).  In the
simulation, it is possible to investigate the relative importance of
bending and twisting in the thread's dynamics by setting the twisting
modulus to zero, taking $2\mu I=0$ in
equation~(\ref{eq:constitutiveBendingTwisting}), while keeping the
bending modulus unchanged.  We performed additional DVT simulations for
a perfectly twist-compliant sewing machine.  The resulting phase diagram shows
only minor differences with Figure~\ref{fig:phasedF}, showing that
twist plays a negligible role, relative to bending, in the selection 
of the stitch patterns.

The experiments and simulations of the viscous sewing machine reveal
in the first instance a great diversity of patterns, whose relations
to one another are not evident.  Our goal was to characterize the
patterns in a more unified way.  This is possible by going beyond a
visual identification and computing for each pattern the Fourier
spectra of the longitudinal and transverse components of the motion of
the thread's contact point with the belt.  We showed that each pattern
has a distinct spectral signature consisting of isolated peaks at a
small number of well-defined frequencies.  The patterns differ from
each other in the values of those frequencies, in their relative
amplitudes, and in their distributions among the longitudinal and
transverse modes.

A closer look shows that the frequencies in the spectra are closely
related to the frequency $\hat\Omega_c$ of steady coiling of a thread
falling on a motionless ($V=0$) surface from the same height.  The
precise nature of the relationship depends on the pattern considered.
For meanders, the frequency of the transverse motion
$\approx\hat\Omega_c$ at onset, but then deviates significantly from
$\hat\Omega_c$ as the belt speed is decreased beyond the critical
value (Figure~\ref{traceOmega}).  In all the other patterns, however,
the frequencies are locked to $\hat\Omega_c$ in some way.  In
stretched coiling, the dominant frequency of both the longitudinal and
transverse motions is $\hat\Omega_c$.  Still more complicated are
double meanders, double coiling, stretched coiling, and the W-pattern,
for which the dominant frequencies are $\hat\Omega_c$ and
$\hat\Omega_c/2$.  The presence of these frequencies reflects the
nonlinear interaction of the two lowest modes of inertio-gravitational
coiling, whose unforced frequencies differ by approximately a factor
of 2 (Figure~\ref{traceOmega}).  Among periodic patterns, the richest
spectral content is achieved by the alternating loops pattern, for
which the five dominant frequencies are multiples of $\hat\Omega_c/3$
(Figure~\ref{traceOmega}).

We proposed a simple kinematic model whereby each pattern is
reconstructed by a superposition of a few frequencies, with
appropriate amplitudes and relative phases in the longitudinal and
transverse directions.  This makes it possible to set up automated
recognition of the patterns, and leads a classification of the
patterns within a unified descriptive framework.  The next step is to
elucidate the physical mechanisms responsible for these simple
spectral signatures.  In future work we plan to investigate the
similarities between the sewing machine and low-dimensional oscillator
models with non-linear forcing, using the direct DVT simulations as a 
starting point to look into these analogies.

We are very grateful to Mikl\`os Bergou and Eitan Grinspun as the
collaborative development of the DVT method has made the present work
possible.  Preliminary numerical results concerning the viscous sewing
machine
have been obtained in collaboration with them~\cite{%
Bergou-Discrete-Geometric-Dynamics-2010,%
Bergou-Audoly-EtAl-Discrete-Viscous-Threads-2010}.  We would like to
thank E.\ Grinspun for suggestions on the manuscript.  During the
preparation of this manuscript, we learnt of an independent but
related experimental work by R.~L.~Welch, B.~Szeto, and S.~W.~Morris\cite{Morris:arXiv}.


\begin{thebibliography}{22}%
\makeatletter
\providecommand \@ifxundefined [1]{%
 \@ifx{#1\undefined}
}%
\providecommand \@ifnum [1]{%
 \ifnum #1\expandafter \@firstoftwo
 \else \expandafter \@secondoftwo
 \fi
}%
\providecommand \@ifx [1]{%
 \ifx #1\expandafter \@firstoftwo
 \else \expandafter \@secondoftwo
 \fi
}%
\providecommand \natexlab [1]{#1}%
\providecommand \enquote  [1]{``#1''}%
\providecommand \bibnamefont  [1]{#1}%
\providecommand \bibfnamefont [1]{#1}%
\providecommand \citenamefont [1]{#1}%
\providecommand \href@noop [0]{\@secondoftwo}%
\providecommand \href [0]{\begingroup \@sanitize@url \@href}%
\providecommand \@href[1]{\@@startlink{#1}\@@href}%
\providecommand \@@href[1]{\endgroup#1\@@endlink}%
\providecommand \@sanitize@url [0]{\catcode `\\12\catcode `\$12\catcode
  `\&12\catcode `\#12\catcode `\^12\catcode `\_12\catcode `\%12\relax}%
\providecommand \@@startlink[1]{}%
\providecommand \@@endlink[0]{}%
\providecommand \url  [0]{\begingroup\@sanitize@url \@url }%
\providecommand \@url [1]{\endgroup\@href {#1}{\urlprefix }}%
\providecommand \urlprefix  [0]{URL }%
\providecommand \Eprint [0]{\href }%
\@ifxundefined \urlstyle {%
  \providecommand \doi  [0]{\begingroup \@sanitize@url \@doi}%
  \providecommand \@doi [1]{\endgroup \@@startlink {\doibase
  #1}doi:\discretionary {}{}{}#1\@@endlink }%
}{%
  \providecommand \doi  [0]{doi:\discretionary{}{}{}\begingroup
  \urlstyle{rm}\Url }%
}%
\providecommand \doibase [0]{http://dx.doi.org/}%
\providecommand \Doi [0]{\begingroup \@sanitize@url \@Doi }%
\providecommand \@Doi  [1]{\endgroup\@@startlink{\doibase#1}\@@Doi}%
\providecommand \@@Doi [1]{#1\@@endlink}%
\providecommand \selectlanguage [0]{\@gobble}%
\providecommand \bibinfo  [0]{\@secondoftwo}%
\providecommand \bibfield  [0]{\@secondoftwo}%
\providecommand \translation [1]{[#1]}%
\providecommand \BibitemOpen [0]{}%
\providecommand \bibitemStop [0]{}%
\providecommand \bibitemNoStop [0]{.\EOS\space}%
\providecommand \EOS [0]{\spacefactor3000\relax}%
\providecommand \BibitemShut  [1]{\csname bibitem#1\endcsname}%
\bibitem [{\citenamefont {Rayleigh}(1914)}]{Rayleigh:1914p5032}%
  \BibitemOpen
  \bibfield  {author} {\bibinfo {author} {\bibfnamefont {L.}~\bibnamefont
  {Rayleigh}},\ }\bibfield  {title} {\enquote {\bibinfo {title} {On the theory
  of long waves and bores},}\ }\href@noop {} {\bibfield  {journal} {\bibinfo
  {journal} {Proceedings of the Royal Society of London. Series A}} (\bibinfo
  {year} {1914})}\BibitemShut {NoStop}%
\bibitem [{\citenamefont {Watson}(1964)}]{Watson:1964p4931}%
  \BibitemOpen
  \bibfield  {author} {\bibinfo {author} {\bibfnamefont {E.}~\bibnamefont
  {Watson}},\ }\bibfield  {title} {\enquote {\bibinfo {title} {The radial
  spread of a liquid jet over a horizontal plane},}\ }\href@noop {} {\bibfield
  {journal} {\bibinfo  {journal} {Journal of Fluid Mechanics}} (\bibinfo {year}
  {1964})}\BibitemShut {NoStop}%
\bibitem [{\citenamefont {Ellegaard}\ \emph {et~al.}(1998)\citenamefont
  {Ellegaard}, \citenamefont {Hansen}, \citenamefont {Haaning}, \citenamefont
  {Hansen}, \citenamefont {Marcussen}, \citenamefont {Bor}, \citenamefont
  {Hansen},\ and\ \citenamefont {Watanabe}}]{ellegaard98}%
  \BibitemOpen
  \bibfield  {author} {\bibinfo {author} {\bibfnamefont {C.}~\bibnamefont
  {Ellegaard}}, \bibinfo {author} {\bibfnamefont {A.~E.}\ \bibnamefont
  {Hansen}}, \bibinfo {author} {\bibfnamefont {A.}~\bibnamefont {Haaning}},
  \bibinfo {author} {\bibfnamefont {K.}~\bibnamefont {Hansen}}, \bibinfo
  {author} {\bibfnamefont {A.}~\bibnamefont {Marcussen}}, \bibinfo {author}
  {\bibfnamefont {T.}~\bibnamefont {Bor}}, \bibinfo {author} {\bibfnamefont
  {J.~L.}\ \bibnamefont {Hansen}}, \ and\ \bibinfo {author} {\bibfnamefont
  {S.}~\bibnamefont {Watanabe}},\ }\bibfield  {title} {\enquote {\bibinfo
  {title} {Creating corners in kitchen sink flows},}\ }\href@noop {} {\bibfield
   {journal} {\bibinfo  {journal} {Nature},\ }\textbf {\bibinfo {volume}
  {392}},\ \bibinfo {pages} {767--768} (\bibinfo {year} {1998})}\BibitemShut
  {NoStop}%
\bibitem [{\citenamefont {Barnes}\ and\ \citenamefont
  {Woodcock}(1958)}]{barnes58}%
  \BibitemOpen
  \bibfield  {author} {\bibinfo {author} {\bibfnamefont {G.}~\bibnamefont
  {Barnes}}\ and\ \bibinfo {author} {\bibfnamefont {R.}~\bibnamefont
  {Woodcock}},\ }\bibfield  {title} {\enquote {\bibinfo {title} {Liquid
  rope-coil effect},}\ }\href@noop {} {\bibfield  {journal} {\bibinfo
  {journal} {Am. J. Physics},\ }\textbf {\bibinfo {volume} {26}},\ \bibinfo
  {pages} {205--209} (\bibinfo {year} {1958})}\BibitemShut {NoStop}%
\bibitem [{\citenamefont {Habibi}\ \emph {et~al.}(2010)\citenamefont {Habibi},
  \citenamefont {Rahmani}, \citenamefont {Bonn},\ and\ \citenamefont
  {Ribe}}]{habibi10}%
  \BibitemOpen
  \bibfield  {author} {\bibinfo {author} {\bibfnamefont {M.}~\bibnamefont
  {Habibi}}, \bibinfo {author} {\bibfnamefont {Y.}~\bibnamefont {Rahmani}},
  \bibinfo {author} {\bibfnamefont {D.}~\bibnamefont {Bonn}}, \ and\ \bibinfo
  {author} {\bibfnamefont {N.~M.}\ \bibnamefont {Ribe}},\ }\bibfield  {title}
  {\enquote {\bibinfo {title} {Buckling of liquid columns},}\ }\href@noop {}
  {\bibfield  {journal} {\bibinfo  {journal} {Phys. Rev. Lett.},\ }\textbf
  {\bibinfo {volume} {104}},\ \bibinfo {pages} {074301} (\bibinfo {year}
  {2010})}\BibitemShut {NoStop}%
\bibitem [{\citenamefont {Habibi}\ \emph {et~al.}(2008)\citenamefont {Habibi},
  \citenamefont {M$\o$ller}, \citenamefont {Ribe},\ and\ \citenamefont
  {Bonn}}]{habibi08}%
  \BibitemOpen
  \bibfield  {author} {\bibinfo {author} {\bibfnamefont {M.}~\bibnamefont
  {Habibi}}, \bibinfo {author} {\bibfnamefont {P.~C.~F.}\ \bibnamefont
  {M$\o$ller}}, \bibinfo {author} {\bibfnamefont {N.~M.}\ \bibnamefont {Ribe}},
  \ and\ \bibinfo {author} {\bibfnamefont {D.}~\bibnamefont {Bonn}},\
  }\bibfield  {title} {\enquote {\bibinfo {title} {Spontaneous generation of
  spiral waves by a hydrodynamic instability},}\ }\href@noop {} {\bibfield
  {journal} {\bibinfo  {journal} {Europhys. Lett.},\ }\textbf {\bibinfo
  {volume} {81}},\ \bibinfo {pages} {38004} (\bibinfo {year}
  {2008})}\BibitemShut {NoStop}%
\bibitem [{\citenamefont {Kaye}(1963)}]{kaye63}%
  \BibitemOpen
  \bibfield  {author} {\bibinfo {author} {\bibfnamefont {A.}~\bibnamefont
  {Kaye}},\ }\bibfield  {title} {\enquote {\bibinfo {title} {A bouncing liquid
  stream},}\ }\href@noop {} {\bibfield  {journal} {\bibinfo  {journal}
  {Nature},\ }\textbf {\bibinfo {volume} {197}},\ \bibinfo {pages} {1001--1002}
  (\bibinfo {year} {1963})}\BibitemShut {NoStop}%
\bibitem [{\citenamefont {Herczynski}\ \emph {et~al.}(2011)\citenamefont
  {Herczynski}, \citenamefont {Cernuschi},\ and\ \citenamefont
  {Mahadevan}}]{Maha-paint}%
  \BibitemOpen
  \bibfield  {author} {\bibinfo {author} {\bibfnamefont {A.}~\bibnamefont
  {Herczynski}}, \bibinfo {author} {\bibfnamefont {C.}~\bibnamefont
  {Cernuschi}}, \ and\ \bibinfo {author} {\bibfnamefont {L.}~\bibnamefont
  {Mahadevan}},\ }\bibfield  {title} {\enquote {\bibinfo {title} {Painting with
  drops, jets, and sheets},}\ }\href@noop {} {\bibfield  {journal} {\bibinfo
  {journal} {Physics Today},\ }\textbf {\bibinfo {volume} {31}},\ \bibinfo
  {pages} {32--36} (\bibinfo {year} {2011})}\BibitemShut {NoStop}%
\bibitem [{\citenamefont {Chiu-Webster}\ and\ \citenamefont
  {Lister}(2006)}]{ChiuWebster:2006p1764}%
  \BibitemOpen
  \bibfield  {author} {\bibinfo {author} {\bibfnamefont {S.}~\bibnamefont
  {Chiu-Webster}}\ and\ \bibinfo {author} {\bibfnamefont {J.}~\bibnamefont
  {Lister}},\ }\bibfield  {title} {\enquote {\bibinfo {title} {The fall of a
  viscous thread onto a moving surface: a 'fluid-mechanical sewing machine'},}\
  }\href@noop {} {\bibfield  {journal} {\bibinfo  {journal} {Journal of Fluid
  Mechanics}} (\bibinfo {year} {2006})}\BibitemShut {NoStop}%
\bibitem [{\citenamefont {Taylor}(1968)}]{taylor68}%
  \BibitemOpen
  \bibfield  {author} {\bibinfo {author} {\bibfnamefont {G.~I.}\ \bibnamefont
  {Taylor}},\ }\bibfield  {title} {\enquote {\bibinfo {title} {Instability of
  jets, threads and sheets of viscous fluid},}\ }in\ \href@noop {} {
  {\bibinfo {booktitle} {Proceedings of the 12th International Congress of
  Applied Mechanics}}}\ (\bibinfo {year} {1968})\BibitemShut {NoStop}%
\bibitem [{\citenamefont {Ribe}\ \emph
  {et~al.}(2006){\natexlab{a}}\citenamefont {Ribe}, \citenamefont {Lister},\
  and\ \citenamefont {Chiu-Webster}}]{Ribe:2006p692}%
  \BibitemOpen
  \bibfield  {author} {\bibinfo {author} {\bibfnamefont {N.~M.}\ \bibnamefont
  {Ribe}}, \bibinfo {author} {\bibfnamefont {J.~R.}\ \bibnamefont {Lister}}, \
  and\ \bibinfo {author} {\bibfnamefont {S.}~\bibnamefont {Chiu-Webster}},\
  }\bibfield  {title} {\enquote {\bibinfo {title} {Stability of a dragged
  viscous thread: Onset of "stitching" in a fluid-mechanical "sewing
  machine"},}\ }\href@noop {} {\bibfield  {journal} {\bibinfo  {journal}
  {Physics of fluids},\ }\textbf {\bibinfo {volume} {18}},\ \bibinfo {pages}
  {124105} (\bibinfo {year} {2006}{\natexlab{a}})}\BibitemShut {NoStop}%
\bibitem [{\citenamefont {Morris}\ \emph {et~al.}(2008)\citenamefont {Morris},
  \citenamefont {Dawes}, \citenamefont {Ribe},\ and\ \citenamefont
  {Lister}}]{Morris:2008p1754}%
  \BibitemOpen
  \bibfield  {author} {\bibinfo {author} {\bibfnamefont {S.}~\bibfnamefont {W.}~\bibnamefont
  {Morris}}, \bibinfo {author} {\bibfnamefont {J.}~\bibnamefont {Dawes}},
  \bibinfo {author} {\bibfnamefont {N.}~\bibnamefont {Ribe}}, \ and\ \bibinfo
  {author} {\bibfnamefont {J.}~\bibnamefont {Lister}},\ }\bibfield  {title}
  {\enquote {\bibinfo {title} {Meandering instability of a viscous thread},}\
  }\href@noop {} {\bibfield  {journal} {\bibinfo  {journal} {Physical Review
  E}} (\bibinfo {year} {2008})}\BibitemShut {NoStop}%
\bibitem [{\citenamefont {Blount}\ and\ \citenamefont
  {Lister}(2011)}]{blount11}%
  \BibitemOpen
  \bibfield  {author} {\bibinfo {author} {\bibfnamefont {M.~J.}\ \bibnamefont
  {Blount}}\ and\ \bibinfo {author} {\bibfnamefont {J.~R.}\ \bibnamefont
  {Lister}},\ }\bibfield  {title} {\enquote {\bibinfo {title} {The asymptotic
  structure of a slender dragged viscous thread},}\ }\href@noop {} {\bibfield
  {journal} {\bibinfo  {journal} {J. Fluid Mech.},\ }\textbf {\bibinfo {volume}
  {674}},\ \bibinfo {pages} {489--521} (\bibinfo {year} {2011})}\BibitemShut
  {NoStop}%
\bibitem [{\citenamefont {Bergou}\ \emph {et~al.}(2010)\citenamefont {Bergou},
  \citenamefont {Audoly}, \citenamefont {Vouga}, \citenamefont {Wardetzky},\
  and\ \citenamefont
  {Grinspun}}]{Bergou-Audoly-EtAl-Discrete-Viscous-Threads-2010}%
  \BibitemOpen
  \bibfield  {author} {\bibinfo {author} {\bibfnamefont {M.}~\bibnamefont
  {Bergou}}, \bibinfo {author} {\bibfnamefont {B.}~\bibnamefont {Audoly}},
  \bibinfo {author} {\bibfnamefont {E.}~\bibnamefont {Vouga}}, \bibinfo
  {author} {\bibfnamefont {M.}~\bibnamefont {Wardetzky}}, \ and\ \bibinfo
  {author} {\bibfnamefont {E.}~\bibnamefont {Grinspun}},\ }\bibfield  {title}
  {\enquote {\bibinfo {title} {Discrete viscous threads},}\ }\Doi
  {10.1145/1778765.1778853} {\bibfield  {journal} {\bibinfo  {journal}
  {Transactions on Graphics},\ }\textbf {\bibinfo {volume} {29}},\ \bibinfo
  {pages} {116} (\bibinfo {year} {2010})}\BibitemShut {NoStop}%
\bibitem [{\citenamefont {Audoly}\ \emph {et~al.}(2011)\citenamefont {Audoly},
  \citenamefont {Clauvelin}, \citenamefont {Bergou}, \citenamefont {Brun},
  \citenamefont {Grinspun},\ and\ \citenamefont
  {Wardetzky}}]{Audoly-Clauvelin-EtAl-Simulating-the-dynamics-of-thin-2011}%
  \BibitemOpen
  \bibfield  {author} {\bibinfo {author} {\bibfnamefont {B.}~\bibnamefont
  {Audoly}}, \bibinfo {author} {\bibfnamefont {N.}~\bibnamefont {Clauvelin}},
  \bibinfo {author} {\bibfnamefont {M.}~\bibnamefont {Bergou}}, \bibinfo
  {author} {\bibfnamefont {P.-T.}\ \bibnamefont {Brun}}, \bibinfo {author}
  {\bibfnamefont {E.}~\bibnamefont {Grinspun}}, \ and\ \bibinfo {author}
  {\bibfnamefont {M.}~\bibnamefont {Wardetzky}},\ }\href@noop {} {\enquote
  {\bibinfo {title} {Simulating the dynamics of thin viscous threads},}\ }
  (\bibinfo {year} {2011}),\ \bibinfo {note} {Submitted  to the Journal of
  Computational Physics}\BibitemShut {NoStop}%
\bibitem [{\citenamefont {Ribe}(2004)}]{Ribe2004}%
  \BibitemOpen
  \bibfield  {author} {\bibinfo {author} {\bibfnamefont {N.~M.}\ \bibnamefont
  {Ribe}},\ }\bibfield  {title} {\enquote {\bibinfo {title} {Coiling of viscous
  jets},}\ }\href@noop {} {\bibfield  {journal} {\bibinfo  {journal}
  {Proceedings of the Royal Society A: Mathematical, Physical and Engineering
  Sciences},\ }\textbf {\bibinfo {volume} {460}},\ \bibinfo {pages}
  {3223--3239} (\bibinfo {year} {2004})}\BibitemShut {NoStop}%
\bibitem [{\citenamefont {Mahadevan}\ \emph {et~al.}(1998)\citenamefont
  {Mahadevan}, \citenamefont {Ryu},\ and\ \citenamefont
  {Samuel}}]{Mahadevan:1998p1053}%
  \BibitemOpen
  \bibfield  {author} {\bibinfo {author} {\bibfnamefont {L.}~\bibnamefont
  {Mahadevan}}, \bibinfo {author} {\bibfnamefont {W.}~\bibnamefont {Ryu}}, \
  and\ \bibinfo {author} {\bibfnamefont {A.}~\bibnamefont {Samuel}},\
  }\bibfield  {title} {\enquote {\bibinfo {title} {Fluid rope trick
  investigated},}\ }\href@noop {} {\bibfield  {journal} {\bibinfo  {journal}
  {Nature},\ }\textbf {\bibinfo {volume} {392}},\ \bibinfo {pages} {140}
  (\bibinfo {year} {1998})}\BibitemShut {NoStop}%
\bibitem [{\citenamefont {Ribe}\ \emph
  {et~al.}(2006){\natexlab{b}}\citenamefont {Ribe}, \citenamefont {Huppert},
  \citenamefont {Hallworth}, \citenamefont {Habibi},\ and\ \citenamefont
  {Bonn}}]{Ribe:2006p691}%
  \BibitemOpen
  \bibfield  {author} {\bibinfo {author} {\bibfnamefont {N.}~\bibnamefont
  {Ribe}}, \bibinfo {author} {\bibfnamefont {H.}~\bibnamefont {Huppert}},
  \bibinfo {author} {\bibfnamefont {M.}~\bibnamefont {Hallworth}}, \bibinfo
  {author} {\bibfnamefont {M.}~\bibnamefont {Habibi}}, \ and\ \bibinfo {author}
  {\bibfnamefont {D.}~\bibnamefont {Bonn}},\ }\bibfield  {title} {\enquote
  {\bibinfo {title} {Multiple coexisting states of liquid rope coiling},}\
  }\href@noop {} {\bibfield  {journal} {\bibinfo  {journal} {Journal of Fluid
  Mechanics},\ }\textbf {\bibinfo {volume} {555}},\ \bibinfo {pages} {275--297}
  (\bibinfo {year} {2006}{\natexlab{b}})}\BibitemShut {NoStop}%
\bibitem [{\citenamefont
  {Trouton}(1906)}]{Trouton-On-the-coefficient-of-viscous-traction-1906}%
  \BibitemOpen
  \bibfield  {author} {\bibinfo {author} {\bibfnamefont {F.~R.~S.}\
  \bibnamefont {Trouton}},\ }\bibfield  {title} {\enquote {\bibinfo {title} {On
  the coefficient of viscous traction and its relation to that of viscosity},}\
  }\href@noop {} {\bibfield  {journal} {\bibinfo  {journal} {Proceedings of the
  Royal Society of London, A},\ }\textbf {\bibinfo {volume} {77}},\ \bibinfo
  {pages} {426--440} (\bibinfo {year} {1906})}\BibitemShut {NoStop}%
\bibitem [{\citenamefont {Buckmaster}\ \emph {et~al.}(1975)\citenamefont
  {Buckmaster}, \citenamefont {Nachman},\ and\ \citenamefont
  {Ting}}]{Buckmaster-Nachman-EtAl-The-buckling-and-stretching-of-a-viscida-1975}%
  \BibitemOpen
  \bibfield  {author} {\bibinfo {author} {\bibfnamefont {J.~D.}\ \bibnamefont
  {Buckmaster}}, \bibinfo {author} {\bibfnamefont {A.}~\bibnamefont {Nachman}},
  \ and\ \bibinfo {author} {\bibfnamefont {L.}~\bibnamefont {Ting}},\
  }\bibfield  {title} {\enquote {\bibinfo {title} {The buckling and stretching
  of a viscida},}\ }\href@noop {} {\bibfield  {journal} {\bibinfo  {journal}
  {Journal of Fluid Mechanics},\ }\textbf {\bibinfo {volume} {69}},\ \bibinfo
  {pages} {1--20} (\bibinfo {year} {1975})}\BibitemShut {NoStop}%
\bibitem [{\citenamefont
  {Bergou}(2010)}]{Bergou-Discrete-Geometric-Dynamics-2010}%
  \BibitemOpen
  \bibfield  {author} {\bibinfo {author} {\bibfnamefont {M.}~\bibnamefont
  {Bergou}},\ } {\bibinfo {title} {Discrete Geometric Dynamics and
  Artistic Control of Curves and Surfaces}},\ \href@noop {} {Ph.D. thesis},\
  \bibinfo  {school} {Columbia University} (\bibinfo {year} {2010})\BibitemShut
  {NoStop}%
\bibitem [{\citenamefont {Welch}\ \emph {et~al.}(2012)\citenamefont {Welch},
  \citenamefont {Szeto},\ and\ \citenamefont
  {Morris}}]{Morris:arXiv}%
  \BibitemOpen
  \bibfield  {author} {\bibinfo {author} {\bibfnamefont {R.}~\bibfnamefont {L.}~\bibnamefont
  {Welch}}, \bibinfo {author} {\bibfnamefont {B.}~\bibnamefont {Szeto}}, \ and\ \bibinfo
  {author} {\bibfnamefont {S.}~\bibfnamefont {W.}~\bibnamefont
  {Morris}},\ }\bibfield  {title}
  {\enquote {\bibinfo {title} {Frequency structure of the nonlinear instability of a dragged viscous thread},}\
  }\href@noop {} {\bibfield  {journal} {\bibinfo  {journal} {arXiv:1201.0817v1}} (\bibinfo {year} {2012})}\BibitemShut {NoStop}%
\end{thebibliography}
%
%

%

\end{document}